%% file: paper.tex
\documentclass[sigplan,10pt]{acmart}

\usepackage{color}
\usepackage{multirow}
\usepackage{url}
\usepackage{graphicx}
\usepackage{algorithm}
\usepackage[noend]{algpseudocode}
\usepackage{subfigure}
\usepackage{booktabs}

\usepackage{enumitem} 
\usepackage{xspace}
\usepackage{latexsym}
\usepackage{amsmath}
\usepackage{amsfonts}
\usepackage{tikz}
\usepackage{tabularx}
\usepackage{balance}
\usepackage{pifont}
\usepackage{soul}
\usepackage{siunitx}
\usepackage{amsthm}
\usepackage{bbm}

\renewcommand\footnotetextcopyrightpermission[1]{}
\setcopyright{none}
\newcommand{\parabf}[1]{\smallskip\noindent\textbf{#1}}

\newcommand{\paraf}[1]{\noindent\textbf{#1}}
\newcommand{\cut}[1]{}

\newcommand{\sysname}{MuxFlow\xspace}
\newcommand{\company}{CompanyX\xspace}
\newcommand{\bytecuda}{xCUDA\xspace}
\newcommand{\sysprobe}{SysMonitor\xspace}

\newcommand{\revise}[1]{#1}

\usepackage{array}
\newcommand{\PreserveBackslash}[1]{\let\temp=\\#1\let\\=\temp}
\newcolumntype{C}[1]{>{\PreserveBackslash\centering}p{#1}}
\newcolumntype{R}[1]{>{\PreserveBackslash\raggedleft}p{#1}}
\newcolumntype{L}[1]{>{\PreserveBackslash\raggedright}p{#1}}

\acmDOI{}

\acmISBN{}

\acmPrice{}

\begin{document}
\sloppy
\date{}

\title{\sysname: Efficient and Safe GPU Sharing in \\
Large-Scale Production Deep Learning Clusters}

\author{Yihao Zhao\footnotemark[1]}
\affiliation{\institution{Peking University}
\country{}}

\author{Xin Liu\footnotemark[1]}
\affiliation{\institution{ByteDance Inc.}
\country{}}

\author{Shufan Liu}
\affiliation{\institution{ByteDance Inc.}
\country{}}

\author{Xiang Li}
\affiliation{\institution{ByteDance Inc.}
\country{}}

\author{Yibo Zhu}
\affiliation{\institution{ByteDance Inc.}
\country{}}

\author{Gang Huang}
\affiliation{\institution{Peking University}
\country{}}

\author{Xuanzhe Liu}
\affiliation{\institution{Peking University}
\country{}}

\author{Xin Jin}
\affiliation{\institution{Peking University}
\country{}}

\input{sections/abstract}

\settopmatter{printfolios=true}
\maketitle
\pagestyle{plain}

\renewcommand{\thefootnote}{\fnsymbol{footnote}}
\footnotetext[1]{Yihao Zhao and Xin Liu contributed equally.}

\input{sections/introduction}
\input{sections/motivation}
\input{sections/architecture}
\input{sections/design}
\input{sections/implementation}
\input{sections/evaluation}
\input{sections/experience}
\input{sections/related}

\input{sections/conclusion}
\label{lastpage}
\clearpage

\bibliographystyle{ACM-Reference-Format}
\bibliography{paper}

\label{wholepage}

\end{document}

%% file: sections/abstract.tex
\begin{abstract}

Large-scale GPU clusters are widely-used to speed up both latency-critical (online) and best-effort (offline) deep learning (DL) workloads.
However, most DL clusters either dedicate each GPU to one workload or share workloads in time, leading to very low GPU resource utilization.

We present \sysname, the first production cluster system that supports efficient and safe space-sharing for DL workloads.
NVIDIA MPS provides an opportunity to share multiple workloads in space on widely-deployed NVIDIA GPUs, but it cannot guarantee the performance and safety of online workloads.
\sysname introduces a two-level protection mechanism for memory and computation to guarantee the performance of online workloads.
Based on our practical error analysis, we design a mixed error-handling mechanism to guarantee the safety of online workloads.
\sysname further proposes dynamic streaming multiprocessor (SM) allocation and matching-based scheduling to improve the efficiency of offline workloads.
\sysname has been deployed at \company's clusters with more than \textbf{\textit{20,000}} GPUs.
The deployment results indicate that \sysname substantially improves the GPU utilization from $26\%$ to $76\%$, SM activity from $16\%$ to $33\%$, and GPU memory from $42\%$ to $48\%$. 
\end{abstract}

%% file: sections/introduction.tex
\section{Introduction}
\label{sec:intro}

Deep learning (DL) has been widely integrated into intelligent applications and services, such as intelligent recommendation~\cite{covington2016deep,gao2021learning},  autonomous driving~\cite{alcon2020timing, jang2020r}, image recognition~\cite{simonyan2014very,he2016deep}, and machine translation~\cite{vaswani2017attention,gehring2017convolutional}.
Some of them provide real-time inference and have critical latency demand (called \textit{online workloads}).
Meanwhile, other workloads do not have hard latency demand (called \textit{offline workloads}). 
Enterprises usually build large-scale GPU clusters for DL workloads and reserve specific GPUs for online workloads.

Existing efforts in online workload management have significantly improved the serving efficiency~\cite{crankshaw2017clipper,gujarati2020serving,shen2019nexus}.
However, a major limitation is that most methods dedicate the whole GPU to a single workload.
It is reported that an online workload usually cannot fully utilize the expensive GPU resource~\cite{ma2020rammer,han2022microsecond} mainly due to two reasons.
First, the frequency of online requests fluctuates from time to time.
When the request frequency is low, more GPU computing units are idle, leading to a great waste of GPU.
Second, even if the request frequency is high, the batch size of online workloads is usually limited to a small value for latency demand and most kernels need few computing resources.
Thus, the computing units in GPU are still underutilized.

A common idea is to share GPUs among multiple workloads with different latency demands~\cite{xiang2019pipelined,xiao2020antman,han2022microsecond}, i.e., sharing GPUs between online and offline workloads.
Time-sharing and space-sharing are two paradigms for GPU sharing.
Time-sharing~\cite{xiao2020antman,cgpu} assigns time slices to different workloads, but it may degrade the performance of online workloads and cannot improve GPU resource utilization in space.
Space-sharing~\cite{han2022microsecond,mps} is a better way to improve GPU resource utilization.
For widely-deployed NVIDIA GPUs, multi-process service (MPS)~\cite{mps} is a feasible choice due to its efficacy, flexibility, and compatibility with NVIDIA GPUs.

However, MPS brings new challenges to production clusters.
First, the primary goal for production clusters is to guarantee the performance of online workloads, such as real-time recommendation and machine translation.
These workloads have hard latency demands because longer latency may influence the user's experience. 
However, MPS cannot guarantee the performance of online workloads.
Second, MPS has a serious error propagation problem, i.e., when one workload encounters an error, the shared workload may also be influenced.
It is critical to guarantee the safety of shared workloads, especially the online workloads in production clusters.

This paper presents \sysname, a system that supports efficient and safe space-sharing of large-scale GPU clusters for DL workloads in the production environment.
\sysname addresses the above challenges to guarantee the performance and safety of online workloads.
\sysname exploits a two-level protection mechanism to guarantee the performance of online workloads from both the workload level and GPU level.
At the workload level, we propose \bytecuda to constrain the GPU memory and computing power used by offline workloads.
\bytecuda monitors the GPU memory allocation to limit the memory usage of offline workloads, and controls kernel launches to limit the computing power used by offline workloads.
Besides, \bytecuda provides adjustable parameters to control how much the online workloads are influenced.
At the GPU level, \sysname employs the \textit{\sysprobe} to monitor the GPU device status.
The \sysprobe maintains a state machine according to multi-dimensional GPU metrics, and will evict offline workloads if the GPU status can potentially compromise the performance of online workloads.

To guarantee the safety of online workloads, we investigate all propagated errors in our production clusters and propose a mixed error-handling mechanism.
We find that $99\%$ propagated errors are caused by \textit{SIGINT} and \textit{SIGTERM} signals, which are usually used to stop containers in Kubernetes~\cite{k8s}.
\sysname employs a graceful exit mechanism that intercepts related signals and releases CUDA context actively to avoid the propagated error.
For other corner cases, \sysname resets the CUDA context and restarts the workloads.

Furthermore, \sysname improves the efficiency of offline workloads.
\sysname dynamically allocates the computing unit of NVIDIA GPU, i.e., streaming multiprocessor (SM), used by offline workloads.
Our key intuition is that we can set the SM percentage of offline workloads complementary to the SM percentage used by online workloads, with an acceptable slowdown of online workloads.
Besides, we observe that different sharing pairs vary dramatically in the efficiency of offline workloads.
To maximize the efficiency, we formulate the problem as a maximum weighted bipartite matching problem.
\sysname exploits a DL approach to build the bipartite graph and the KM algorithm~\cite{kuhn1955hungarian,munkres1957algorithms} to solve this problem.

In summary, we make the following contributions.

\begin{itemize}[leftmargin=*]
    \item We investigate the characteristics of production inference clusters and identify the opportunity in space sharing to better utilize GPU.
    \item We provide efficient and safe space-sharing with three mechanisms: a two-level protection mechanism to guarantee the performance of online workloads, a mixed error-handling mechanism to ensure the safety of online workloads, and a dynamic SM percentage mechanism to improve the efficiency of offline workloads.
    \item We design a matching-based scheduling algorithm to improve the sharing efficiency at the cluster level.
    The scheduling algorithm can improve the overall normalized throughput of offline workloads while maintaining the performance of online workloads.
    \item We introduce \sysname, the first production cluster system that enables efficient and safe space sharing.
    We have deployed \sysname in a production cluster with more than $20,000$ GPUs at \company to serve tens of thousands of daily workloads.
    Deployment results show that \sysname improves the GPU utilization from $26\%$ to $76\%$, SM activity from $16\%$ to $33\%$, and GPU memory from $42\%$ to $48\%$.
\end{itemize}

%% file: sections/motivation.tex
\section{Motivation}
\label{sec:motivation}

In this section, we begin with introducing DL workloads and critical terminologies.
Then we describe the observations from the production cluster for online workloads to motivate the design of \sysname.
We end by discussing opportunities to share the GPUs between different DL workloads.

\subsection{DL workloads}
DL workloads use the deep neural network (DNN) to perform inference or training.
DL workloads are usually classified into two categories, i.e., online workload and offline workload, according to the latency demand.
\emph{Online workload} refers to latency-critical inference, such as real-time recommendation~\cite{covington2016deep,gao2021learning} and machine translation~\cite{vaswani2017attention,gehring2017convolutional}. 
Online workloads have strict latency demands because longer end-to-end latency may hurt users' experience.
Additionally, the requests for online workloads are usually submitted periodically at different frequencies.
\emph{Offline workload} does not have strong latency demand, such as DL training,  batch inference, scientific computing~\cite{senior2020improved}, and automatic neural architecture search~\cite{liu2018progressive,tan2019mnasnet}.
These workloads usually take hours or even days to finish.
The offline workloads do not have hard time requirements and can usually highly utilize the computing units of GPU, making them suitable to fill the idle GPU resource.

\subsection{Production cluster for online workloads}
Production clusters exploit GPUs to accelerate DL workloads~\cite{crankshaw2017clipper,gujarati2020serving,han2022microsecond}.
GPUs are usually assigned to online workloads exclusively to guarantee the latency demand.
We study GPU resource utilization in production clusters from two aspects: memory and computing power.

\begin{figure}[t]
        \centerline{\includegraphics[width=\linewidth,trim=0 0 0 20,clip]{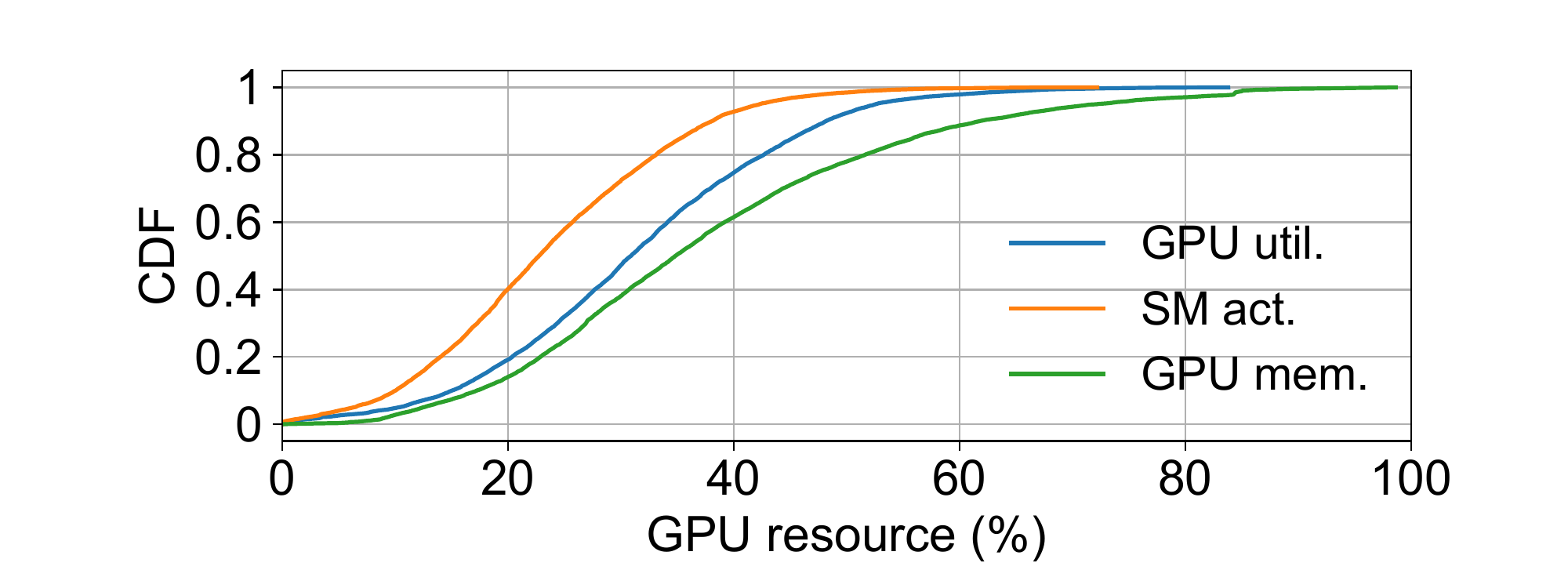}}
        \vspace{-0.2in}
        \caption{GPU resource statistic in a production cluster for online workloads.}
        \vspace{-0.1in}
        \label{fig:motiv_gpu_resource}
\end{figure}
\parabf{Low GPU resource utilization.}
We collect one week's statistics of GPU computation utilization and memory usage in the inference cluster of \company, as shown in Figure~\ref{fig:motiv_gpu_resource}.
\revise{
The inference workloads include various popular DL models, such as CNN, GNN, LLM, and recommendation models.
}
As for the GPU computing utilization, we use two metrics: GPU utilization and SM activity~\cite{dcgm}.
GPU utilization and SM utilization represent how busy the GPU is in time and in space, respectively.
GPU memory usage is the ratio of used memory to memory capacity.
Figure~\ref{fig:motiv_gpu_resource} illustrates that both GPU utilization and SM utilization are lower than $60\%$ for more than $99\%$ GPUs.
In addition, GPU memory usage is less than $60\%$ for about $90\%$ GPUs.
These numbers show that GPUs are underutilized in both memory and computing power, indicating a great waste of valuable GPUs.

\parabf{Fluctuating and predictable GPU utilization.}
\begin{figure}[t]
        \centerline{\includegraphics[width=\linewidth,trim=0 0 0 20,clip]{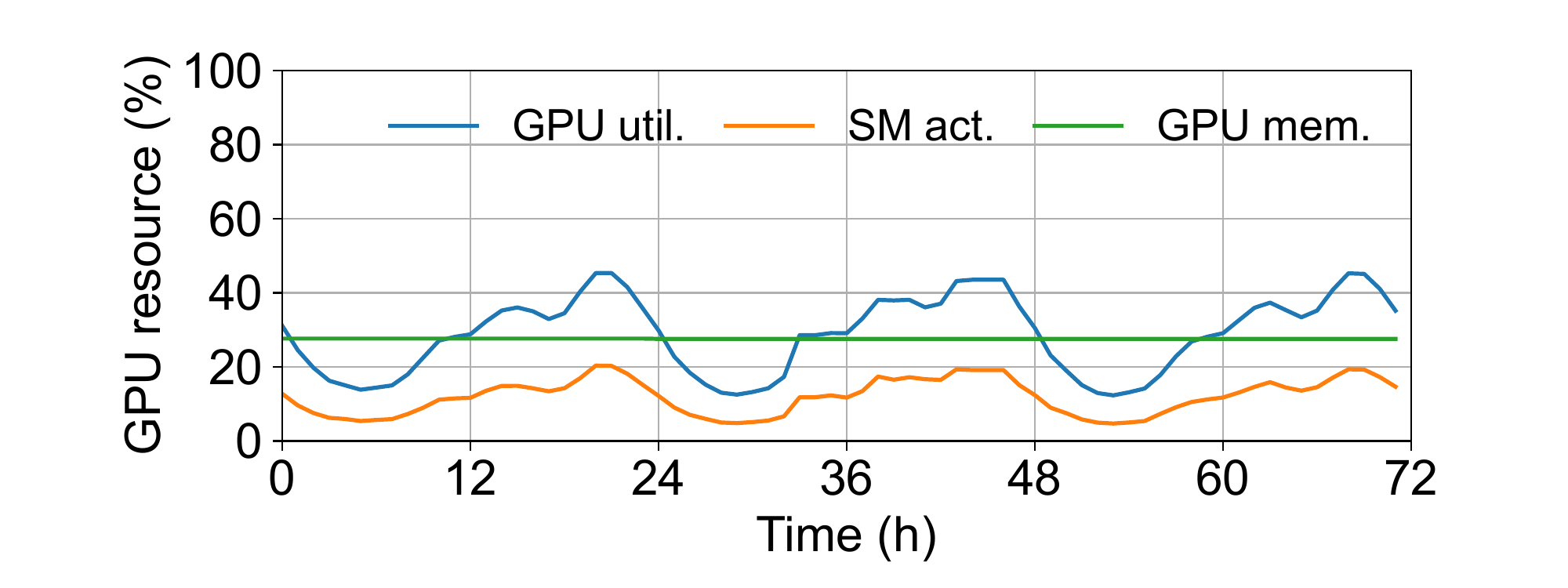}}
        \vspace{-0.15in}
        \caption{Resource usage of one typical online workload. GPU util., SM act., and GPU mem. are short for GPU utilization, SM activity, and GPU memory usage, respectively.}
        \vspace{-0.1in}
        \label{fig:motiv_host}
\end{figure}
We take one typical online workload in the production cluster of \company as an example and show its GPU computing utilization and memory usage in Figure~\ref{fig:motiv_host}.
Both the GPU utilization and SM activity fluctuate greatly in one day, because the number of online requests varies from time to time.
For example, more users use entertainment applications in the evening and send more online requests to related services, while during the day, fewer requests are sent.
The GPU memory usage is stable because the DL framework, e.g., PyTorch~\cite{paszke2019pytorch}, caches the intermediate GPU memory for efficiency.
Besides, we observe that the curves of the GPU usage metrics are smooth in minutes and periodical in days.
Thus, we can predict the GPU usage metrics by the past values.

\begin{figure}[t]
        \centerline{\includegraphics[width=0.9\linewidth]{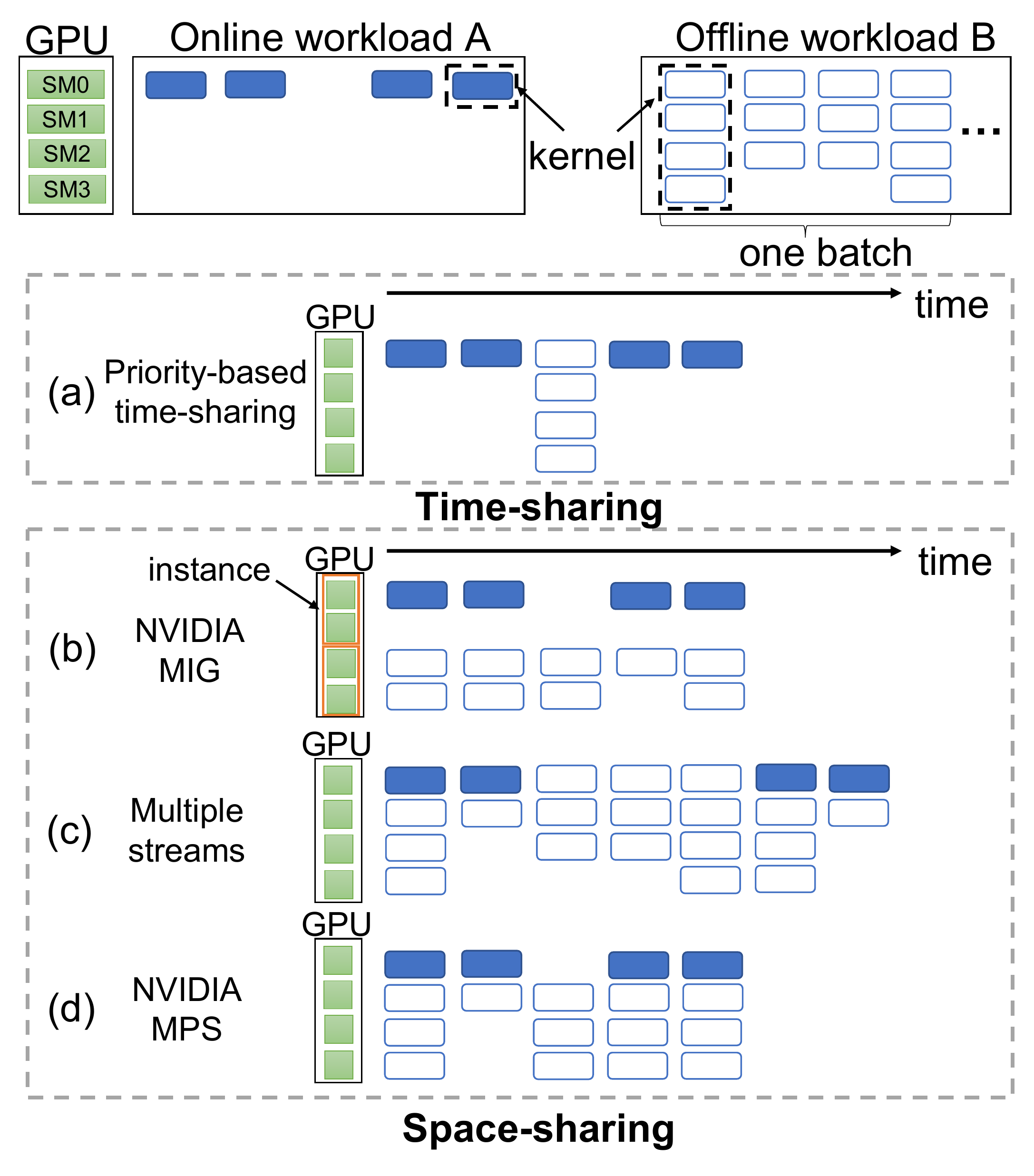}}
        \vspace{-0.15in}
        \caption{An example of different GPU sharing approaches for NVIDIA GPUs.}
        \vspace{-0.1in}
        \label{fig:motiv_gpusharing}
\end{figure}

\subsection{Opportunities in GPU sharing}
Some recent work~\cite{cgpu,mig, multi-streams, mps} has exploited GPU sharing approaches to improve GPU resource utilization. 
There are two aspects of GPU sharing, i.e., time-sharing and space-sharing.
We compare GPU sharing approaches for widely-deployed NVIDIA GPUs with an example as shown in Figure~\ref{fig:motiv_gpusharing}.

\parabf{Time-sharing is not efficient to improve GPU resource utilization.}
In time-sharing, shared workloads use different time slices.
To protect the performance of online workloads, priority-based time-sharing~\cite{xiao2020antman,cgpu} (Figure~\ref{fig:motiv_gpusharing}(a)) assigns more time slices to high-priority workloads.
However, a single online workload usually cannot fill all SMs of one GPU completely~\cite{han2022microsecond, ma2020rammer}, leading to a waste of GPU computing power.

\parabf{Opportunity: space-sharing of GPU.}
When a workload cannot fully utilize the GPU computing units, i.e., SMs, it can share the idle SMs with other workloads.
The SMs of one GPU can be divided into multiple parts, and used by different workloads simultaneously, i.e., space-sharing.
We summarize three space-sharing approaches to share widely-deployed NVIDIA GPUs in Figure~\ref{fig:motiv_gpusharing}.
NVIDIA proposes \textit{multi-instance GPU (MIG)}~\cite{mig} which can partition one GPU into multiple instances, as shown in Figure~\ref{fig:motiv_gpusharing}(b).
However, the partition cannot be dynamically adjusted during workload execution, and thus, we have to allocate maximum resources for online workloads which leads to a waste of GPU.
\revise{
Additionally, MIG only works for specific new-generation GPU types, e.g., A100 and H100, which are not widely used in production clusters.
}
CUDA provides \textit{multiple streams}~\cite{multi-streams} (Figure~\ref{fig:motiv_gpusharing}(c)) to execute kernels from multiple workloads, whereas the concurrent workloads can significantly degrade the performance of online workloads.
Besides, NVIDIA stream can only share with other streams in one process, which needs to merge multiple workloads and is hard to manage in production clusters.
We find that \textit{NVIDIA MPS}~\cite{mps} (Figure~\ref{fig:motiv_gpusharing}(d)) is the best trade-off between GPU resource utilization and online performance.
\revise{
MPS is supported by Kepler and newer NVIDIA GPUs which are the majority of the GPUs used in production clusters.
}
MPS enables NVIDIA GPU to execute multiple workloads at the same time by assigning different sets of SMs to the shared workloads.
Besides, MPS provides environment variables to roughly control the SM percentage used by each workload, which enables performance protection of online workloads.

\begin{figure}[t]
	\subfigure[]{
        \begin{minipage}{0.47\linewidth}
        \centerline{\includegraphics[width=\linewidth,trim=0 0 0 0,clip]{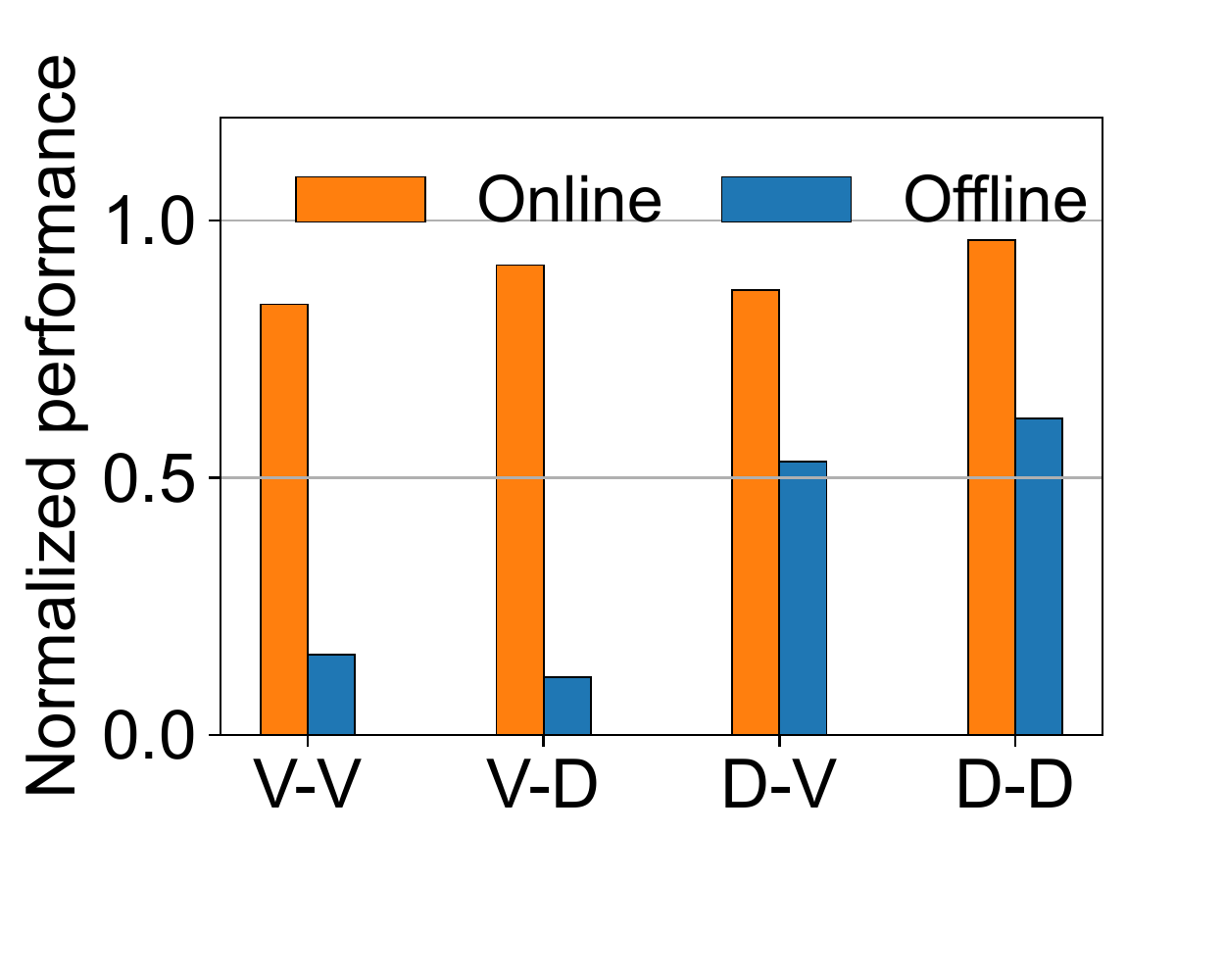}}
        \label{fig:motiv_mps}
        \end{minipage}
        }
	\subfigure[]{
        \begin{minipage}{0.47\linewidth}
        \centerline{\includegraphics[width=\linewidth,trim=0 0 0 0,clip]{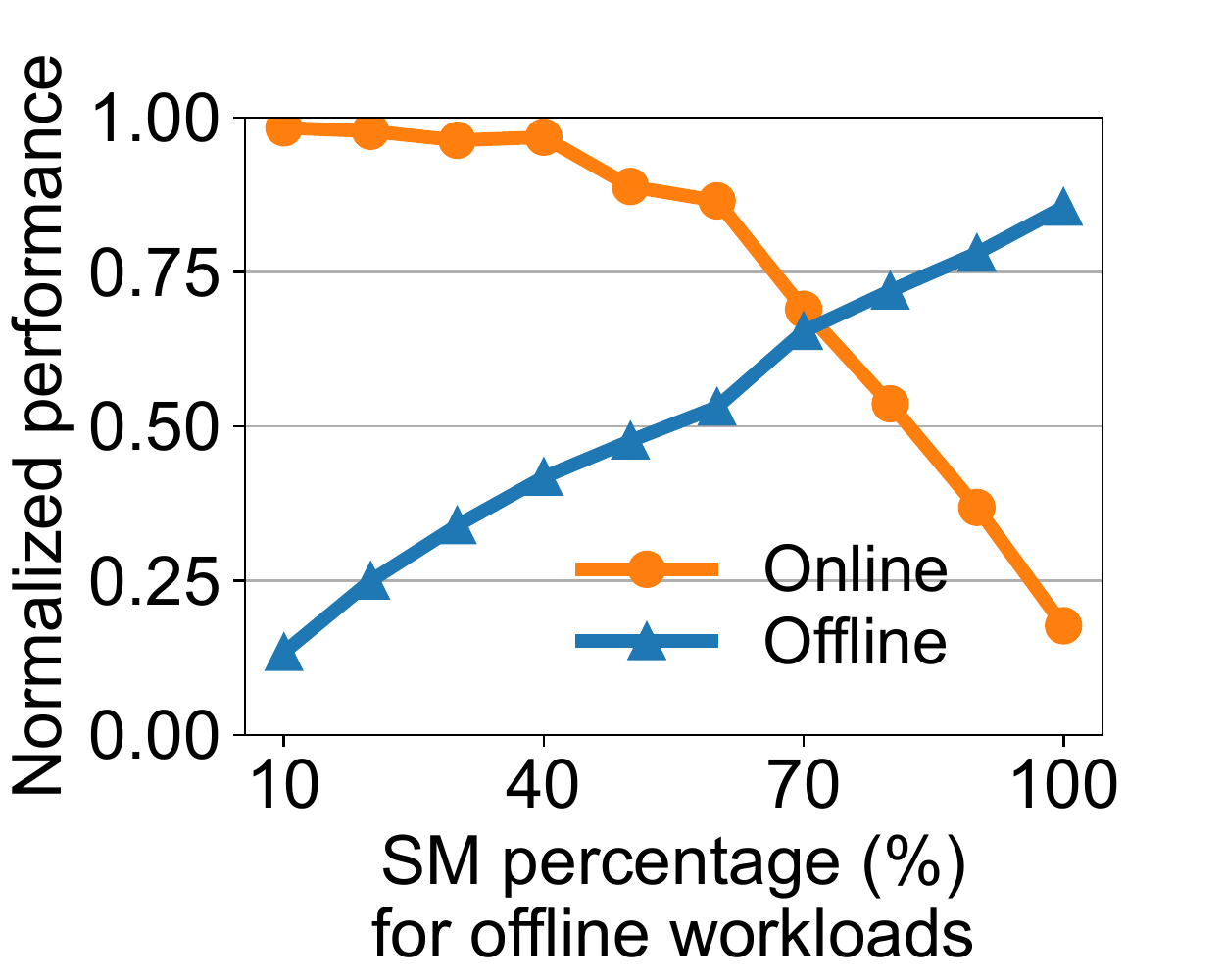}}
        \label{fig:motiv_cmatp}
        \end{minipage}
    }
     \vspace{-0.1in}
     \caption{(a) GPU sharing with MPS and adjusted SM percentage (A-B represents sharing one online workload A with one offline workload B. V, D are short for VGG16 and DenseNet201, respectively), and (b) Impact of the SM percentage for offline workloads (DenseNet201 as the online workload and VGG16 as the offline workload).}
     \vspace{-0.1in}
    \label{fig:motiv_mps_all}
\end{figure}

To show the effect of MPS, we choose two DL models, VGG16~\cite{simonyan2014very} and DenseNet201~\cite{huang2017densely} as workloads.
We use the inference of these DL models as online workloads and the training as offline workloads.
These workloads are tested on NVIDIA T4 GPU.
Figure~\ref{fig:motiv_mps} reports the normalized performance when we share one online workload with one offline workload.
The normalized performance is the average iteration duration when running alone divided by the average iteration duration when shared with other workloads.
To protect the performance of online workloads, we adjust the SM percentage of offline workloads.
Figure~\ref{fig:motiv_mps} demonstrates that one GPU can provide up to $62\%$ more computing power while slowing online workloads less than $20\%$.
The results indicate the potential of sharing GPU in space with MPS.

\parabf{Challenges for space-sharing.}
There are some technical challenges to deploy space-sharing in large-scale DL clusters.
First, the primary goal of production clusters is to guarantee the performance of online workloads.
MPS enables us to roughly change the SM percentage used by each workload.
However, it cannot guarantee the performance of online workloads.
For example, the online requests may suddenly burst due to a special activity, but the SM percentage for offline workloads cannot be reduced timely.
Thus, we need to control the execution process of the shared workloads to protect online workloads.
Second, MPS is notorious for its serious error propagation problems.
Specifically, when one workload encounters an error, the shared workload will also be influenced.
These safety problems are intractable and critical in production clusters.
Third, different SM percentages assigned to offline workloads can greatly influence the efficiency of both shared workloads.
We change the SM percentage assigned to offline workloads from $10\%$ to $100\%$ as shown in Figure~\ref{fig:motiv_cmatp}.
The normalized performance of both online workload and offline workload varies more than $5\times$.
Thus, choosing a proper SM percentage is important to provide efficient GPU sharing.
Fourth, different shared pairs of online and offline workloads show different impacts on the shared workloads in Figure~\ref{fig:motiv_mps}.
The normalized performance of offline workloads varies up to $50\%$ in Figure~\ref{fig:motiv_mps}.
Additionally, the number of possible sharing plans is factorial to the number of workloads, which is enormous for a production cluster.
We need to efficiently decide how to share workloads to maximize offline efficiency, while maintaining the performance of online workloads.

%% file: sections/architecture.tex
\section{\sysname architecture}
\label{sec:arch}

\begin{figure}[t]
        \centerline{\includegraphics[width=\linewidth]{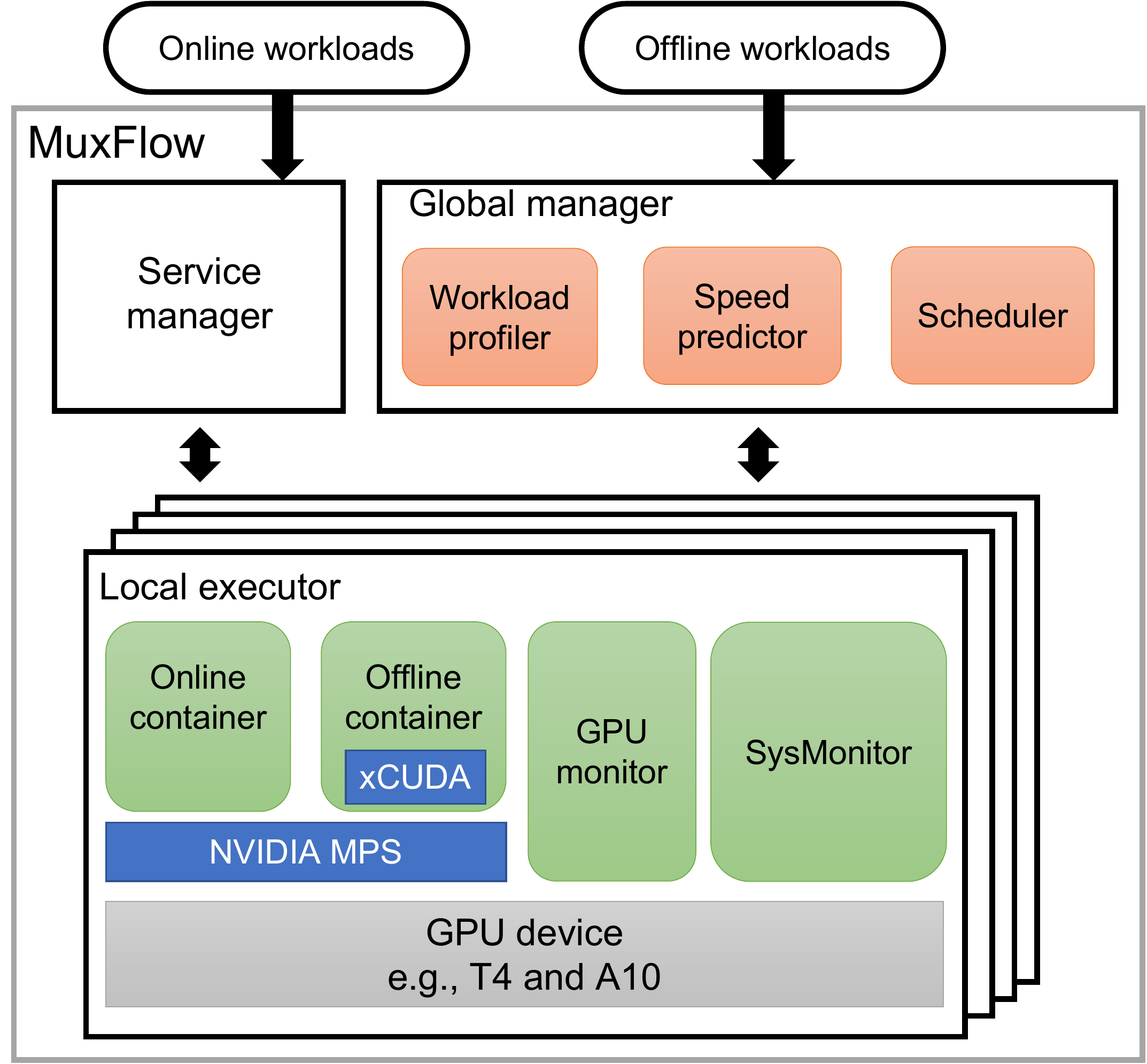}}
        \vspace{-0.1in}
        \caption{\sysname architecture.}
        \vspace{-0.1in}
        \label{fig:arch_arch}
\end{figure}

\sysname is a cluster system that enables efficient and safe space-sharing in production clusters with tens of thousands of heterogeneous GPUs at \company.
The architecture of \sysname is shown in Figure~\ref{fig:arch_arch}.
\sysname consists of a service manager for online workloads, a global manager for offline workloads, and a set of local executors for each GPU.

\parabf{Service manager.}
In this paper, we only describe the functionality of the service manager briefly as the details of the service manager are beyond the scope of this paper.
The service manager is responsible for online workloads deployment, online requests discovery, and horizontal pod autoscaling.

\parabf{Global manager.} 
When an offline workload is submitted to \sysname, the global manager buffers the offline workload in a pending queue and makes scheduling decisions periodically.
The global manager includes three components: workload profiler, speed predictor, and scheduler.

\textit{Workload profiler.}
The workload profiler gets the GPU resource utilization and execution time for each offline workload.
When an offline workload is first submitted, the workload profiler runs the job for a few iterations and measures the execution information.
The measured information is stored in a database and can be used by the speed predictor when making scheduling decisions.

\textit{Speed predictor.}
When the speed predictor gets an online workload and an offline workload, it can predict the sharing speed of the given workloads.
The speed predictor employs a DL model to perform prediction.
The DL model leverages the execution information when the workloads are executed separately.
The execution information is reported by the GPU monitor for online workloads and is profiled by the workload profiler in advance for offline workloads.
The predicted speed is passed to the scheduler to make scheduling decisions.

\textit{Scheduler.}
The scheduler schedules the offline workloads from the pending queue.
By utilizing the predicted speed from the speed predictor, the scheduler exploits a matching-based scheduling algorithm to decide which offline workload and online workload should share the same GPU.
The scheduling algorithm can find the optimal sharing strategy.
The scheduler performs global rescheduling periodically at a fixed interval.

\parabf{Local executor.}
Each local executor manages the workloads on one GPU.
The local executor executes workloads according to the scheduling decision of the scheduler and monitors the running workloads.
Besides, it can evict the offline workload if the online workload is under threat.
The workloads share the same GPU in space with MPS.
There are four components in the local executor: online container, offline container with \bytecuda, GPU monitor, and \sysprobe.

\textit{Online container and offline container.}
The online container runs the server for online workload and serves the online requests from upstream callers registered in the server manager.
The offline container runs the offline workload.
With \bytecuda built in the offline container, the execution of the offline workload is controlled to guarantee the performance of the online workload.
To do so, \bytecuda limits the GPU memory and computing power used by the offline workloads.

\textit{GPU monitor.}
The GPU monitor periodically collects GPU metrics, such as GPU utilization, memory usage, and SM clock.
These data reflect the workload pressure of each GPU and they are leveraged by the \sysprobe and \bytecuda to manage offline workloads.

\textit{\sysprobe.}
The \sysprobe maintains a state machine reflecting the device status and ensures that the device is not in unhealthy status.
The state machine transits according to the GPU metrics collected by the GPU monitor.
When the state machine indicates one online workload is highly influenced, the \sysprobe will evict the shared offline workload. 

%% file: sections/design.tex
\section{Efficient and safe space-sharing}
\label{sec:space-sharing}
The goal of \sysname is to guarantee the performance and safety of online workloads, and improve the efficiency of offline workloads.
In this section, we introduce how to provide efficient and safe space-sharing in each local executor.
We first describe how we protect the performance of online workloads with a two-level protection mechanism.
Then we introduce how we protect the safety of online workloads with a mixed error-handling mechanism.
We end with the dynamic SM allocation mechanism for offline efficiency improvement.

\subsection{Two-level performance protection}
\begin{figure*}[t!]
	\subfigure[\bytecuda]{
        \begin{minipage}{0.6\linewidth}
        \centerline{\includegraphics[width=0.9\linewidth]{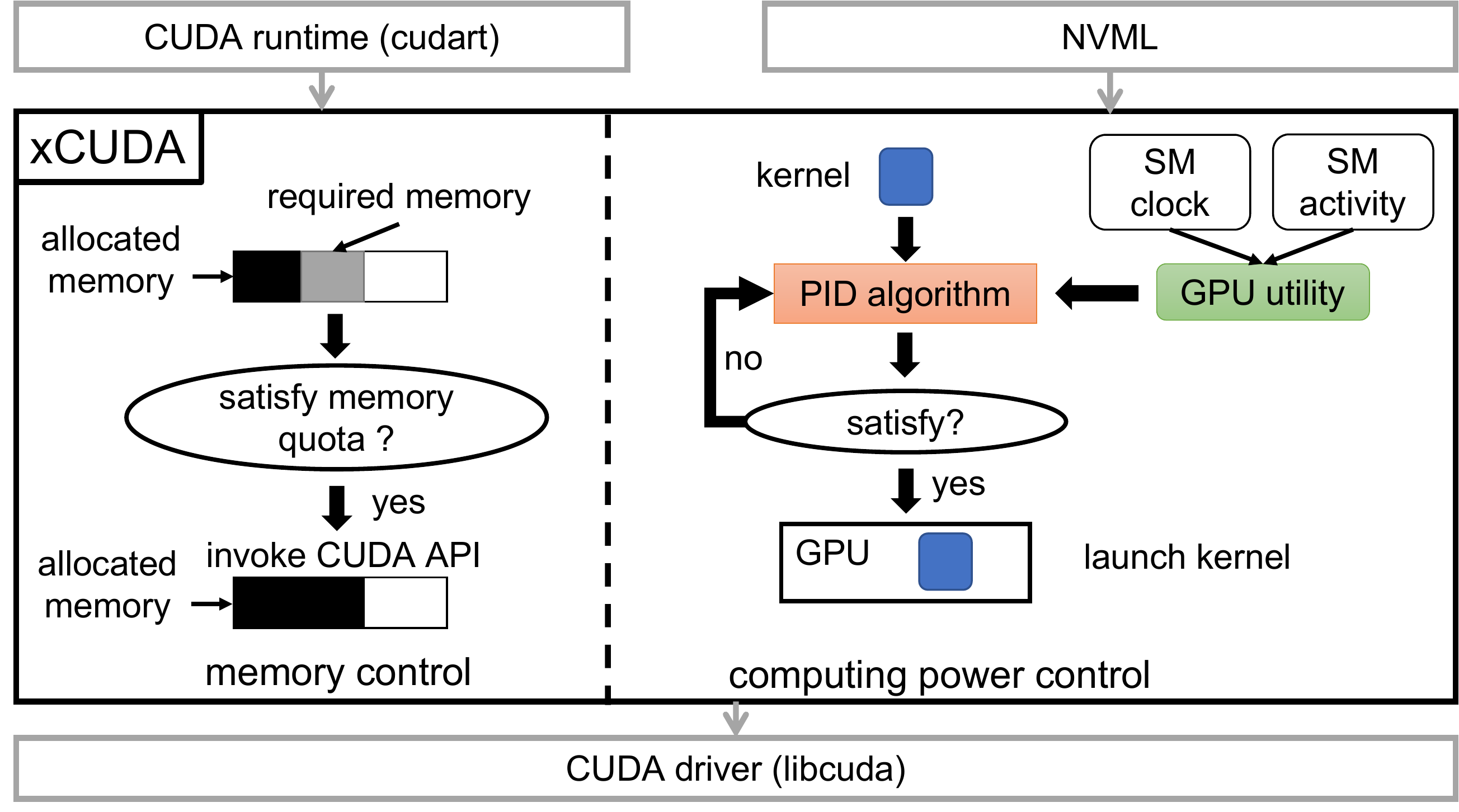}}
        \label{fig:design_bytecuda}
        \end{minipage}
    }
	\subfigure[The state machine of \sysprobe]{
    \begin{minipage}{0.3\linewidth}
        \centerline{\includegraphics[width=\linewidth]{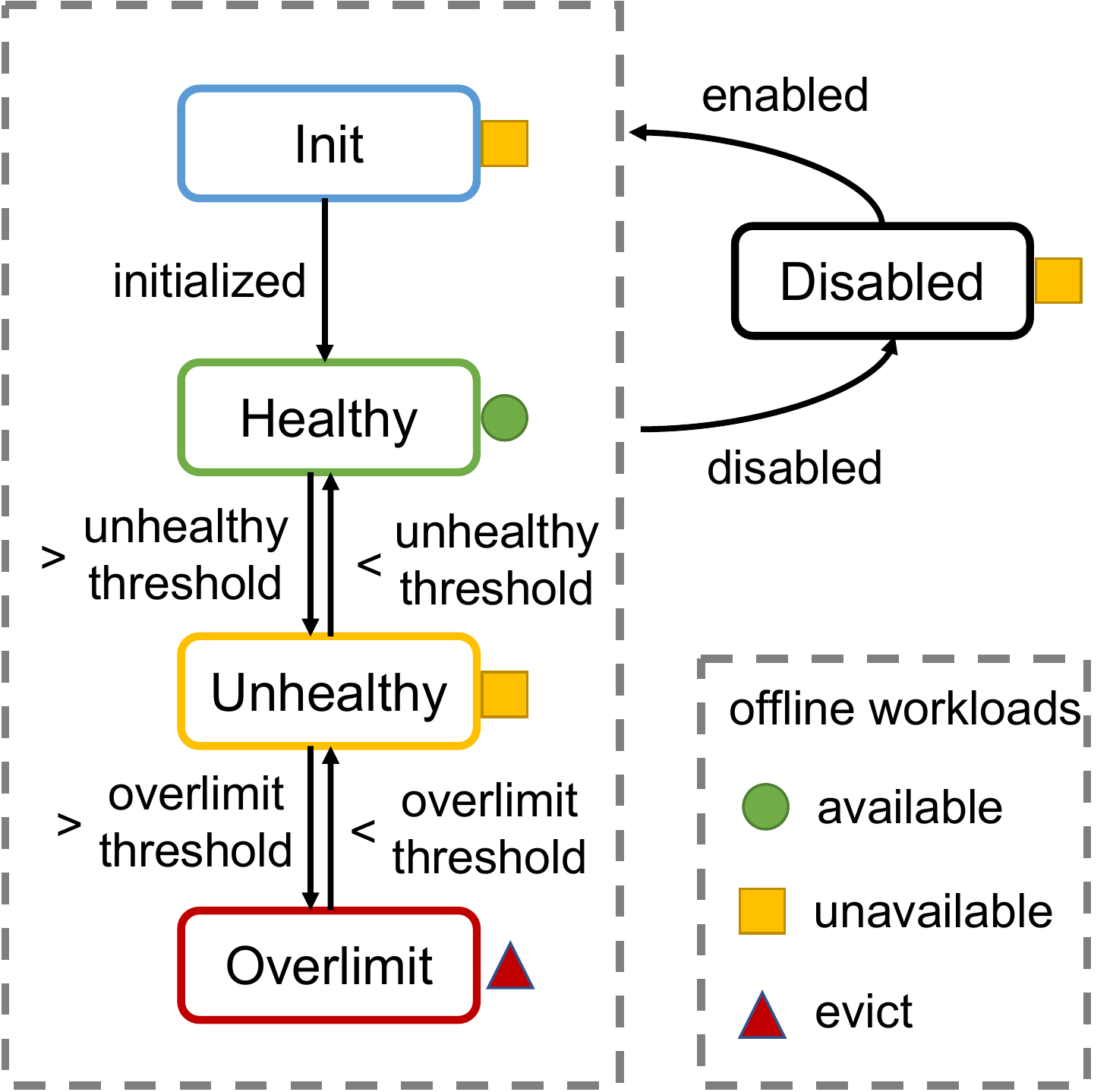}}
        \label{fig:design_sysprobe}
     \end{minipage}}
     \vspace{-0.2in}
     \caption{Two-level performance protection for online workloads. \bytecuda and \sysprobe protect the performance of online workloads from workload level and GPU level, respectively.}
     \vspace{-0.1in}
    \label{fig:design_two-level}
\end{figure*}

MPS provides an environment variable to control the SM percentage used by a workload.\footnote{$CUDA\_MPS\_ACTIVE\_THREAD\_PERCENTAGE$ configures the active thread percentage at the client process level and limits the SM percentage used by the client process.} 
We can roughly limit the offline workload with this environment variable and reduce the slowdown of online workloads indirectly.
However, in production clusters, we need rigorous protection mechanisms for online workloads.
\sysname employs a two-level protection mechanism as shown in Figure~\ref{fig:design_two-level}.
Specifically, \sysname controls offline workloads to protect online workloads at the workload level by \bytecuda and the GPU level by \sysprobe.

First of all, we need GPU metrics to make decisions on how to control the offline workloads.
In the local executor, the GPU monitor collects real-time GPU metrics periodically.
The collection interval is in the millisecond level for timely control.
The metrics include GPU resource utilization (e.g., GPU utilization, SM activity, and GPU memory usage), and GPU device status (e.g., SM clocks, power, and temperature).
The GPU monitor stores the metrics for only several minutes because old data not only consume storage but also are useless for timely workload management.

\parabf{Workload level.} 
\bytecuda is built in the offline container to control the GPU memory and computing power used by offline workloads, as shown in Figure~\ref{fig:design_bytecuda}.
In the aspect of memory, \bytecuda can keep track of the GPU memory usage and make sure that the memory used by the offline workload does not exceed the GPU memory quota.
Specifically, whenever the DL framework, e.g., TensorFlow~\cite{abadi2016tensorflow} and PyTorch~\cite{paszke2019pytorch}, applies for GPU memory by calling the related CUDA API, the call will be checked by \bytecuda first.

In terms of computing power, we aim to guarantee the performance of online workloads and improve GPU computing utilization.
\revise{
For NVIDIA GPUs, the SM clock represents how fast the SMs execute instructions.
The performance of online workloads is greatly affected by the SM clock, and the SM clock will decrease when the GPU load is high.
The decrease in SM clock is especially noteworthy for NVIDIA GPU for inference, e.g., T4.
}
Thus, our goal is equivalent to attaining both high SM clock and high GPU utilization.
When the SM clock is low, we can delay the kernel launches of the offline workloads to reduce the GPU load and improve the SM clock.
When the GPU utilization is low, we can launch more kernels to improve it.
Formally, we define $U_{SM}$ as the SM activity and $a_C$ as a clock factor that is negatively correlated with the SM clock.
We use the GPU load $U_{GPU}$ to quantify our goal, which can be calculated by,
\begin{equation}
\label{equ:gpu_load}
U_{GPU} = U_{SM} \times a_C .
\end{equation}

\revise{
However, the SM clock and GPU utilization are conflicting in practice because the SM clock is negatively correlated to the GPU load, while the GPU utilization is positively correlated to the GPU load.
Note that when sharing online and offline workloads, it is enough to get an SM clock which is similar to the SM clock when the online workload runs separately.
Consequently, \bytecuda sets an SM clock threshold for these two goals.
When the SM clock is below the threshold, \bytecuda tends to improve the SM clock.
When the SM clock is over the threshold, \bytecuda tends to improve the GPU utilization.
The factor $a_C$ can be calculated by,
}
\begin{equation}
a_C =
\begin{cases}
1 + a_L*\frac{T_{SM}-C_{SM}}{T_{SM}} & C_{SM} < T_{SM} \\ 
1 - a_H*\frac{C_{SM}-T_{SM}}{C_{H}-T_{SM}} & C_{SM} \ge T_{SM} ,
\end{cases}
\label{equ:clock_factor}
\end{equation}
where $a_L$ is a parameter for low SM clock, $a_H$ is a parameter for high SM clock, $C_{SM}$ is the SM clock, $T_{SM}$ is a SM clock threshold, and $C_{H}$ is the highest SM clock.
\revise{
$a_L$ is much larger than $a_H$ to show the preference of increasing the SM clock when it is below the threshold.
}
With the GPU load $U_{GPU}$, \bytecuda will delay the kernel when $U_{GPU}$ is high and launch the kernel when $U_{GPU}$ is low.
Additionally, as the GPU load $U_{GPU}$ may change rapidly, \bytecuda leverages the PID algorithm~\cite{johnson2005pid} to provide more stable and robust controlling.

\parabf{GPU level.}
\bytecuda constrains offline workloads and provides indirect performance protection for online workloads.
However, it cannot reply to changes caused by online workloads in time.
For example, when the GPU memory usage of the online workload bursts, \bytecuda cannot dynamically adjust the GPU memory quota for offline workloads.
Thus, at the GPU level, \sysname uses \sysprobe to monitor the GPU device status with a state machine.
Figure~\ref{fig:design_sysprobe} shows the state machine of \sysprobe.
The state machine has five states and each state has a set of metric thresholds for GPU utilization, SM activity, SM clock, and GPU memory usage.
The threshold values are empirically selected.
\revise{
Note that offline workloads can only be scheduled to \textit{Healthy} GPUs.
}

The five states of \sysprobe are as follows:
(1) \textit{Init state} represents that the GPU is being initialized. When the initialization finishes, the Init state will transit to the Healthy state.
(2)\textit{Healthy state} represents that the GPU is healthy and is able to execute offline workloads. 
The metric thresholds of this state guarantee that the online workload is not influenced by the offline workloads.
Once one metric reaches the Unhealthy threshold, the state will transit to the Unhealthy state.
Furthermore, once one metric exceeds the Overlimit threshold, the state will directly transit to the Overlimit state.
(3)\textit{Unhealthy state} represents that one GPU metric is in the Unhealthy state and none is in the Overlimit state.
Intuitively, this state means that the online workloads may be influenced.
The offline workloads are forbidden to be scheduled to the GPU in this state.
Once one metric exceeds the Overlimit threshold, the state will transit to the Overlimit state.
Oppositely, if all metrics are below the Healthy threshold, the state will transit to the Healthy state.
(4) \textit{Overlimit state} represents that the GPU device is overloaded.
In this state, the offline workloads are evicted.
When all metrics are below the Overlimit threshold after a period, the state will transit to the Unhealthy state.
To avoid frequent eviction, the period is set to be the exponent of the times going into the Overlimit state during the last two hours.
(5)\textit{Disabled state} represents that the GPU device is unavailable and no workload runs on it.

\begin{figure}[t]
        \centerline{\includegraphics[width=\linewidth]{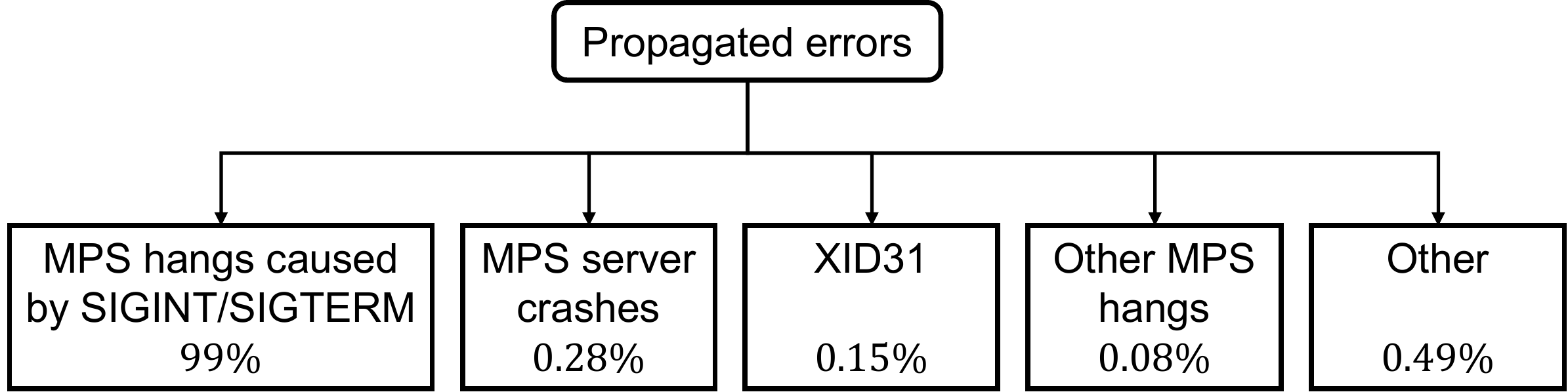}}
        \vspace{-0.15in}
        \caption{Propagated errors in a production cluster.}
        \vspace{-0.1in}
        \label{fig:design_error}
\end{figure}

\subsection{Safety protection}
MPS has a serious error propagation problem, i.e., one workload's error may impact other workloads sharing the same GPU.
The error propagation problem is dangerous in large-scale clusters, especially for online workloads.
For example, if one offline workload is canceled by SIGINT signal, the MPS context may hang and the shared online workload cannot serve requests.
We summarize propagated errors in one production cluster with MPS enabled as shown in Figure~\ref{fig:design_error}, and propose a mixed error-handling mechanism.

We find that $99\%$ of such propagated errors are caused by SIGINT/SIGTERM.
To handle the dominant error type, we use \bytecuda to intercept SIGINT and SIGTERM signals and exit gracefully.
Specifically, when \bytecuda gets these signals, it will freeze all kernel launches and release CUDA context actively.
Other errors only count to $1\%$, e.g., MPS server crash (caused by program bugs), XID31 event (GPU memory page fault), and MPS hangs caused by other reasons.
For these errors, we summarize their error patterns.
An automated detector monitors GPUs and alerts when the error patterns are satisfied.
Once \bytecuda gets the alert, it will reset the CUDA context and MPS server.

\begin{figure*}[t]
	\subfigure[Online workloads only]{
        \begin{minipage}{0.3\linewidth}
        \centerline{\includegraphics[width=0.75\linewidth]{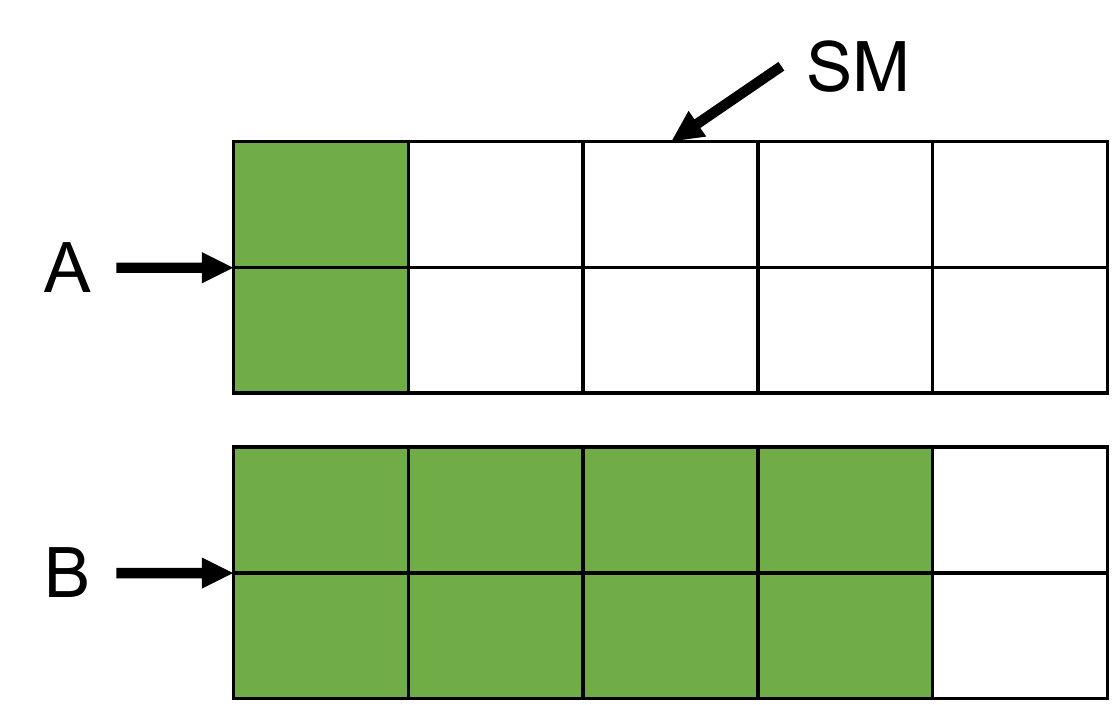}}
        \label{fig:design_dynamic_sm_0}
        \end{minipage}
    }
	\subfigure[Sharing with fixed SM percentage]{
        \begin{minipage}{0.3\linewidth}
        \centerline{\includegraphics[width=0.9\linewidth]{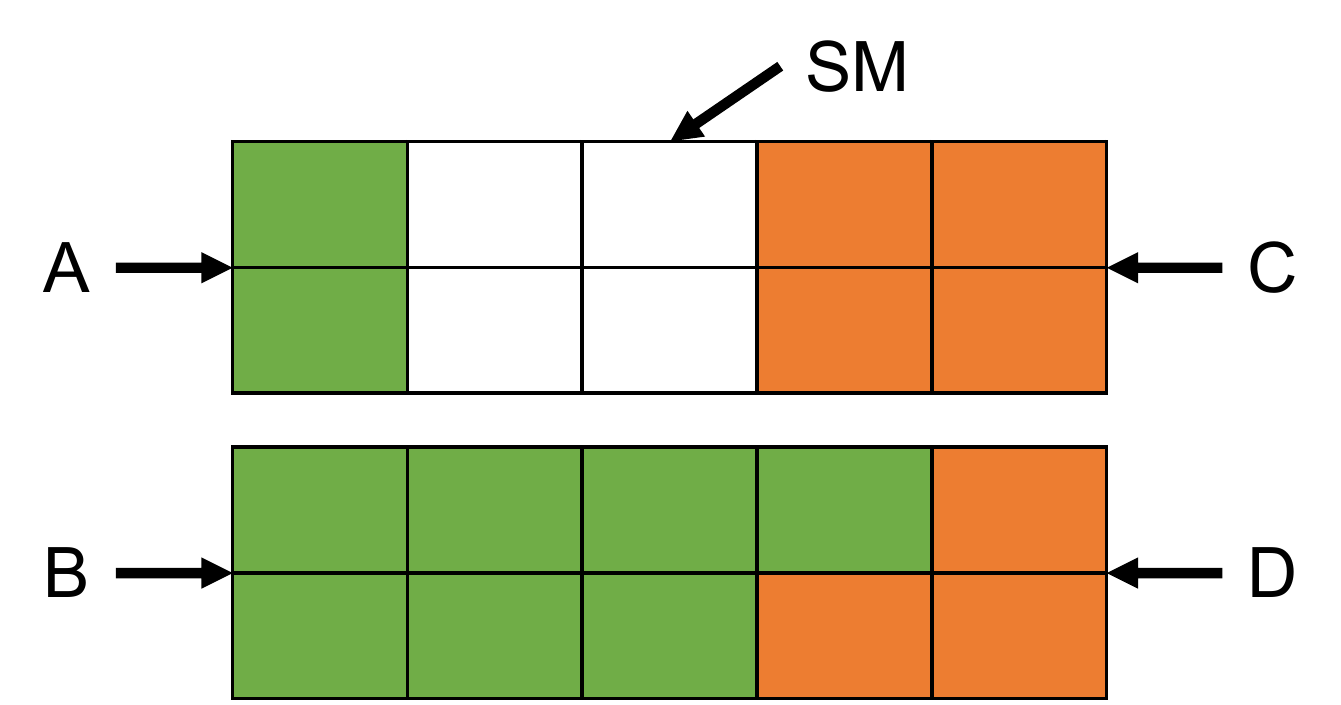}}
        \label{fig:design_dynamic_sm_1}
        \end{minipage}
    }
	\subfigure[Sharing with dynamic SM percentage]{
        \begin{minipage}{0.3\linewidth}
        \centerline{\includegraphics[width=0.9\linewidth]{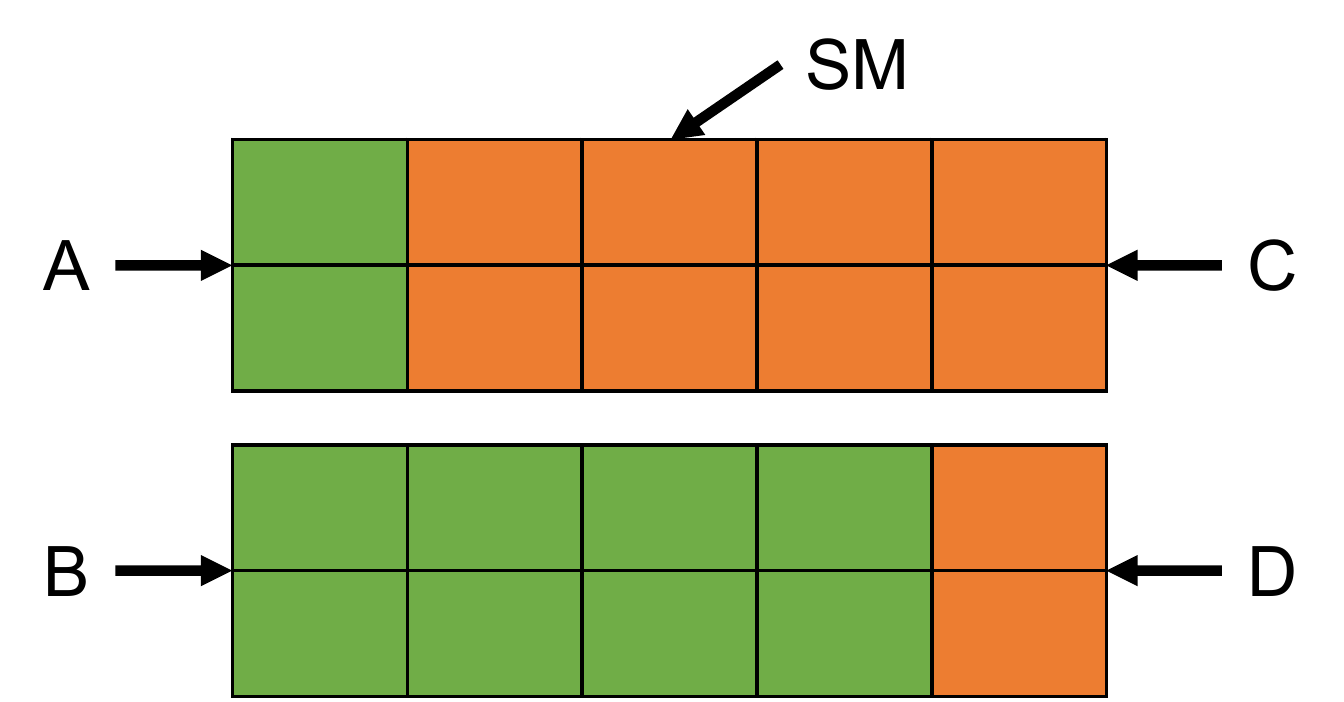}}
        \label{fig:design_dynamic_sm_2}
        \end{minipage}
    }
     \vspace{-0.15in}
     \caption{Dynamic SM allocation. A and B are two online workloads. C and D are two offline workloads. (b) The SM percentages for offline workloads are fixed at $40\%$. (c) The SM percentages for offline workloads are dynamically adjusted. }
     \vspace{-0.1in}
    \label{fig:design_dynamic_sm}
\end{figure*}

\subsection{Dynamic SM allocation}
In Figure~\ref{fig:motiv_cmatp}, we illustrate that the SM percentage assigned to offline workloads can influence the speed of shared workloads dramatically.
In other words, we can balance the speed of the online workload and that of the offline workload by changing the SM allocation.
Our goal is to maximize the efficiency of offline workloads with an acceptable slowdown of the online workloads.
As SM is the computing unit of GPU, maximizing the efficiency of offline workloads is approximately equal to maximizing the percentage of SMs assigned to offline workloads.
Apparently, fixed SM allocation is not a panacea for all sharing cases.
We use the example in Figure~\ref{fig:design_dynamic_sm} to illustrate the drawbacks of fixed SM allocation.
A and B are online workloads, and C and D are offline workloads.
Assume that the SM percentage is set to $40\%$ for offline workloads.
The online workload A only uses $20\%$ SMs and leaves more than $40\%$ SMs.
If we fix the SM percentage for offline workloads to $40\%$, there will be $40\%$ idle SM and the computing power is wasted.
The online workload B uses $80\%$ SMs when running alone and the left SMs are less $40\%$.
With the fixed SM allocation, the offline workload D will occupy B's SM and slow the online workload B down.

To provide efficient space-sharing, we propose the dynamic SM allocation mechanism by selecting the proper SM percentage for offline workloads.
A natural idea is to assign the SM percentage according to the SM activity of online workloads, as shown in Figure~\ref{fig:design_dynamic_sm_2}.
For example, we can set the SM percentage for the offline workload C to $80\%$ and D to $20\%$.
In this way, the computing units, i.e., SMs, are used up and the shared workloads do not contend for SMs.

\begin{figure}[t]
        \centerline{\includegraphics[width=\linewidth]{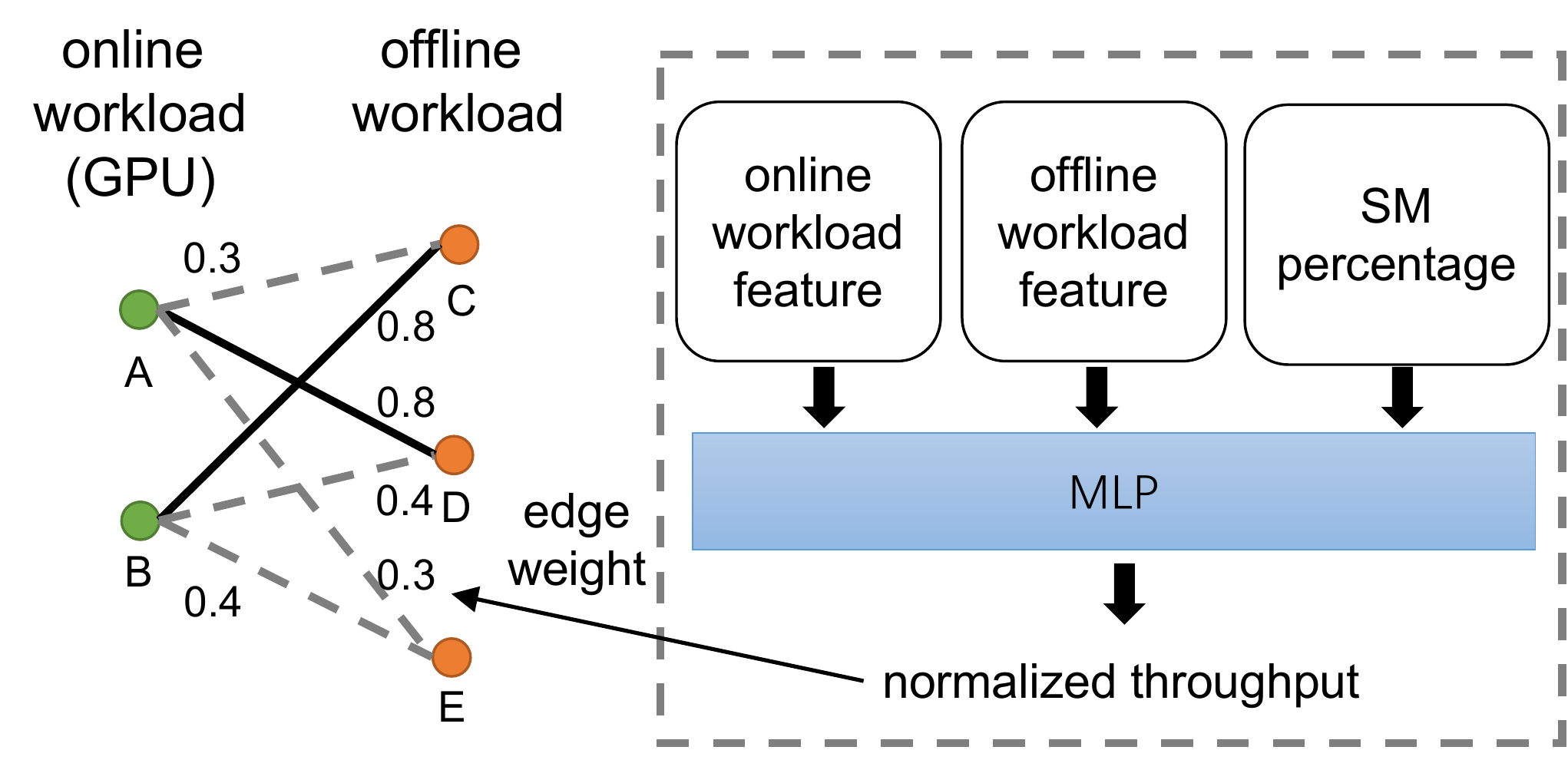}}
        \vspace{-0.1in}
        \caption{Scheduling plans are computed with maximum weighted bipartite matching. The edge weights represent the normalized throughput of offline workloads. }
        \vspace{-0.1in}
        \label{fig:design_matching}
\end{figure}

\section{Matching-based scheduling}
\label{sec:schedule}

Note that one offline workload has diverse throughput when sharing with different online workloads.
We next consider scheduling submitted workloads to achieve high overall throughput for offline workloads.
The method of scheduling and deploying online workloads is orthogonal to the scheduling algorithm of offline workloads.
For online workloads, we reuse the scheduling and deployment strategy of the exclusive inference cluster at \company.
The details of the strategy are beyond the scope of this paper.
For offline workloads, the scheduling algorithm needs to solve two problems.

The first problem is to capture the overall throughput for all offline workloads.
It is unfair and meaningless to simply sum up the throughput of every offline workload because different kinds of workloads vary in throughput when running separately.
Thus, we use the normalized throughput, which is defined as the throughput when sharing divided by the throughput when running separately.
We can get the throughput of separate execution by profiling the offline workloads when it is submitted.
However, the shared throughput is difficult to get because a production cluster usually has thousands of online workloads and it is impossible to profile all sharing pairs.
Fortunately, getting the shared throughput can be seen as a regression problem, which can be solved by DL.
We can use the profiled separate execution features of online and offline workloads as input, and employ a DL model to get shared throughput.
Specifically, we choose highly related execution features, e.g., GPU utilization, SM activity, SM occupancy, separate execution time, and assigned SM percentage, as input.
We employ the multi-layer perceptron (MLP) as the speed predictor to get the shared throughput.
Because different GPU types perform diversely, we train multiple MLPs for each GPU type.

The second problem is to select the best sharing plan from all potential sharing plans.
For a production cluster with thousands of online and offline workloads, there is an enormous number of combinations to share workloads.
Formally, given $n$ online workloads and $m$ offline workloads, we pair the workloads to maximize the normalized throughput of the entire sharing plan.
This problem can be transformed to a maximum weighted bipartite matching problem.
Figure~\ref{fig:design_matching} shows how we model this problem.
We build a bipartite graph $G(V, E)$, where each node $v\in V$ represents a workload.
Every node belongs to one of the two node sets, i.e., the online workload set and the offline workload set.
The edge $(u,v)\in E$ represents sharing online workload $u$ and offline workload $v$ with the normalized throughput of $v$ as the edge weight.
A matching of $G$ is a set of disjoint edges and it corresponds to a feasible sharing plan.
Finding the sharing plan with maximum overall throughput is converted to finding the maximum weighted matching of the corresponding bipartite graph.
We leverage the Kuhn-Munkres (KM) algorithm~\cite{kuhn1955hungarian,munkres1957algorithms} for this well-studied problem.
The KM algorithm can find the optimal maximum weighted bipartite matching in $O(|V|^3)$.

\revise{
We use an example in Figure~\ref{fig:design_matching} to illustrate the matching problem.
There are two online workloads (A and B) and three offline workloads (C, D, and E).
The number on the edge represents the normalized throughput of the offline workload when sharing with the online workload.
For example, when sharing C with A, the normalized throughput of C is $0.3$.
We compare two matching plans.
Plan 1 shares A with D and B with C.
The overall throughput of offline workloads is $0.8+0.8=1.6$.
Plan 2 shares A with C and B with E.
The overall throughput of offline workloads is $0.3+0.4=0.7$.
Apparently, plan 1 has higher overall throughput and is more efficient in the efficiency of offline workloads.
}

Algorithm~\ref{alg:scheduling} shows the pseudocode of the matching-based scheduling.
Given online workloads $W_{on}$ and offline workloads $W_{off}$, we initialize the bipartite graph $G$ where online workloads $W_{on}$ and offline workloads $W_{off}$ are two disjoint sets of $G$ (Line 1-4).
For each pair of nodes, we get the SM percentage for the offline workload by dynamic SM allocation mechanism (Line 5-6).
The edge weights are calculated by the speed predictor $P$ (Line 7-8).
Then we get the maximum weighted bipartite matching $M$ by the KM algorithm (Line 9-10).
Edge $(u,v)$ in matching $M$ represents that the offline workload $v$ should be shared with the online workload $u$.

\begin{algorithm}[t]
\caption{Scheduling algorithm of \sysname}
\begin{algorithmic}[1]
\begin{small}
\Statex \textbf{Input:} Online workloads $W_{on}$; offline workloads $W_{off}$; speed predictor $P$.

\Statex \textbf{Main routine:}
\begin{algorithmic}[1]
    \State \textbf{// Initialize}
    \State $G.Init()$
    \State $G.AddNodes(W_{on})$
    \State $G.AddNodes(W_{off})$
    \For{each pair $(u,v)$, where $u\in W_{on}, v\in W_{off}$}
        \State $sm \gets DynamicSM(u,v)$
        \State $weight \gets P.CalcNormTput(u, v, sm)$
        \State $G.AddEdge(u,v,weight)$
    \EndFor
    \State \textbf{// Find optimal matching with the KM algorithm}
    \State $M \gets G.GetMatching()$
\end{algorithmic}

\Statex \textbf{Subroutines:}
\Statex \textbullet~$G.Init()$: Initialize an empty bipartite graph.
\Statex \textbullet~$G.AddNodes(W)$: Add every workload $w\in W$ as a node to graph $G$.
\Statex \textbullet~$G.AddEdge(u,v,c)$: Add an edge $(u,v)$ with edge weight $c$ to graph $G$.
\Statex \textbullet~$G.GetMatching()$: Calculate the maximum weighted bipartite matching of graph $G$ by the KM algorithm.
\Statex \textbullet~$DynamicSM(u,v)$: Use dynamic SM allocation mechanism to get the proper SM percentage for offline workload $v$.
\Statex \textbullet~$P.CalcNormTput(u,v,sm)$: Calculate the normalized throughput of $v$ by the speed predictor $P$.

\end{small}
\end{algorithmic}
\label{alg:scheduling}
\end{algorithm}

%% file: sections/implementation.tex
\section{Implementation}
\label{sec:impl}

At \company, we have deployed \sysname in our internal clusters to serve daily DL workloads.
The internal clusters consist of heterogeneous GPUs, including NVIDIA T4 GPU and NVIDIA A10 GPU.
Integrated with Kubernetes~\cite{k8s}, \sysname manages thousands of GPUs in each cluster and more than 20,000 GPUs in all.

\parabf{Service manager.}
For online workloads, we use the existing service manager at \company which deploys containers, discovers service, and autoscales horizontal pods.

\parabf{Global manager.}
We modify the Kubernetes scheduler to schedule offline workloads.
The workload profiler takes several dedicated GPUs, whose number is negligible to the total number of GPUs.
When a new offline workload comes, the workload profiler performs a few dry runs of the workload and utilizes the NVIDIA Data Center GPU Manager (DCGM) tools~\cite{dcgm} and NVIDIA Management Library (NVML)~\cite{nvml} libraries to collect GPU metrics.
We collect about 2,000 data for each GPU type to train the speed predictor.
The MLPs of the speed predictor have four layers with hidden size $64\times 64$.
The MLPs are trained with momentum SGD optimizer~\cite{ruder2016overview} in PyTorch v1.8.0~\cite{paszke2019pytorch} until they converge.
\sysname invokes the scheduler periodically to schedule all offline workloads.
When moving workloads, we record checkpoints of offline workloads and restart the workloads after transmitting the models and checkpoints.
As the datasets are usually colossal, we store the datasets in a remote file system and fetch data during the execution.
We implement the scheduler as a third-party plugin to the Kubernetes scheduler.

\parabf{Local executor.}
Each local executor executes online workloads according to the service manager and offline workloads according to the global manager.
DL workloads are executed in Docker containers with our customized components.
We add Best-Effort GPU DevicePlugin in Kubernetes and relevant control paths with Kubelet and \sysprobe for offline workloads.
To control SM percentage, we leverage the environment variable $CUDA\_MPS\_ACTIVE\_THREAD\_PERCENTAGE$ provided by MPS.
The GPU monitor collects resource metrics through DCGM~\cite{dcgm} and NVML~\cite{nvml} for NVIDIA GPU.
The \sysprobe updates the state machine with the collected resource metrics and empirically-set thresholds.
When the state is unhealthy, the \sysprobe will ask the NodeManager in Kubernetes to evict offline workloads.
\bytecuda intercepts nearly 800 CUDA driver APIs for GPU memory allocation and kernel launch.
The GPU memory quota of offline workloads is fixed to $40\%$ as Figure~\ref{fig:motiv_gpu_resource} reports that most online workloads use less than $60\%$ GPU memory.
We adopt the cpuset of Cgroup for CPU isolation.
For memory, \sysname will evict offline workloads if memory usage is higher than a threshold or the kernel swap daemon is busy for a long time.
The parameters to calculate GPU load in Equation~\ref{equ:gpu_load}$\&$\ref{equ:clock_factor} are empirically selected through trial-and-error.

%% file: sections/evaluation.tex
\section{Evaluation}
\label{sec:evaluation}

Our evaluation consists of testbed experiments, trace-driven simulations, and results from the production deployment.
We mainly focus on the efficiency and safety of \sysname.
The results show that \sysname oversells up to $90.0\%$ GPU resources to offline workloads, and improves the GPU utilization by $4.0\times$, SM activity by $4.7\times$, and GPU memory usage by $1.5\times$.
The error rate of \sysname is similar to the dedicated inference clusters in production deployment at \company.

\subsection{Experimental setup}
\paraf{Testbed.}
We conduct the testbed experiments with $125$ machines and $1,000$ GPUs.
Each machine is equipped with $8$ NVIDIA Tesla T4 GPUs, 2 Intel Xeon Platinum CPUs, 128G memory, and 100+25G NIC.
We PyTorch v1.8.0 with CUDA 11.1 for offline workloads.
We use real online workloads in our production clusters and these workloads use different DL frameworks and CUDA drivers.

\parabf{Simulator.}
Inspired by ~\cite{gu2019tiresias,zhao2022multi}, we build a simulator to evaluate a broader set of configurations, traces, and baselines.
We profile the iteration duration, GPU resource utilization, and device metrics of the selected DL workloads when they are executed separately and shared with others.
The profiling results include more than 200 executions.
The difference between the simulation and testbed experiment is under $5\%$, showing the high fidelity of the simulator.

\parabf{Workloads.}
We use the actual online workloads and real-time requests at \company for the testbed experiment.
\revise{
The online workloads use a wide spectrum of DL models, including CNN, GNN, LLM, and recommendation models.
}
For trace-driven simulation, we select three online workloads deployed over hundreds of GPUs at \company, and generate requests according to their actual query per second (QPS) varying from 20 to 190.
For offline workloads, we leverage the public trace from Microsoft~\cite{jeon2019analysis} and split the trace according to the virtual cluster ID.
We use submission time and duration from the traces and randomly choose DL models from four popular DL models including ResNet50~\cite{he2016deep}, VGG16~\cite{simonyan2014very}, DensNet201~\cite{huang2017densely}, and Inception-V3~\cite{szegedy2016rethinking}, in accordance with the common practice~\cite{gu2019tiresias,zhao2022multi,han2022microsecond}.
We repeat the workloads to fit in 1,000 GPUs and guarantee that the generated traces can be finished in 12 hours for the testbed experiment and 24 hours for simulations.
The numbers of offline workloads in these traces vary from 1,410 to 7,287.
We set the batch size according to the memory quota limited by \bytecuda.

\parabf{Baselines.}
We compare \sysname with three related systems, Online-only, Time-sharing, and Priority-based time-sharing (PB-time-sharing).
Online-only executes only the online workloads and shows the optimal latency for online workloads.
Time-sharing shares workloads in time and assigns the time slices of GPUs to the shared workloads by the GPU driver strategy, which is adopted by Gandiva~\cite{xiao2018gandiva}.
PB-time-sharing sets online workloads with high priority and assigns more time slices of GPUs to high-priority workloads to protect the high-priority workloads' performance, which is adopted by AntMan~\cite{xiao2020antman} and PAI~\cite{weng2022pai}.

\parabf{Metrics.}
Average latency and $99\%-th$ latency are two common metrics to evaluate the performance of online workloads~\cite{gujarati2020serving,han2022microsecond}.
Average job completion time (JCT) and makespan are used to reflect the workload efficiency of schedulers~\cite{xiao2020antman,zhao2022multi}.
Offline normalized throughput shows the sharing efficiency.
\revise{
We define how much GPU the offline workloads get in the aspect of computation speed as the oversold GPU.
This metric is in the range of $[0,1]$, where $0$ represents that the offline workloads get no GPU computation resource, and $1$ represents that the offline workloads get equivalent GPU computation resources as they are executed exclusively.}
This metric can be calculated by Equation~\ref{equ:oversold}.
\begin{equation}
Oversold\ GPU=\frac{\sum_{w\in W_{off}} T^{real}_w}{\sum_{w\in W_{off}} T^{sep}_w} ,
\label{equ:oversold}
\end{equation}
where $W_{off}$ represents offline workloads, $T^{real}_w$ represents the real execution time of $w$, and $T^{sep}_w$ represents the execution time of $w$ when running exclusively.
We report GPU resource utilization with three metrics: GPU utilization, SM activity, and GPU memory utilization.

\subsection{Testbed experiments}

\begin{figure}[t]
        \centerline{\includegraphics[width=0.95\linewidth,trim=0 0 0 50, clip]{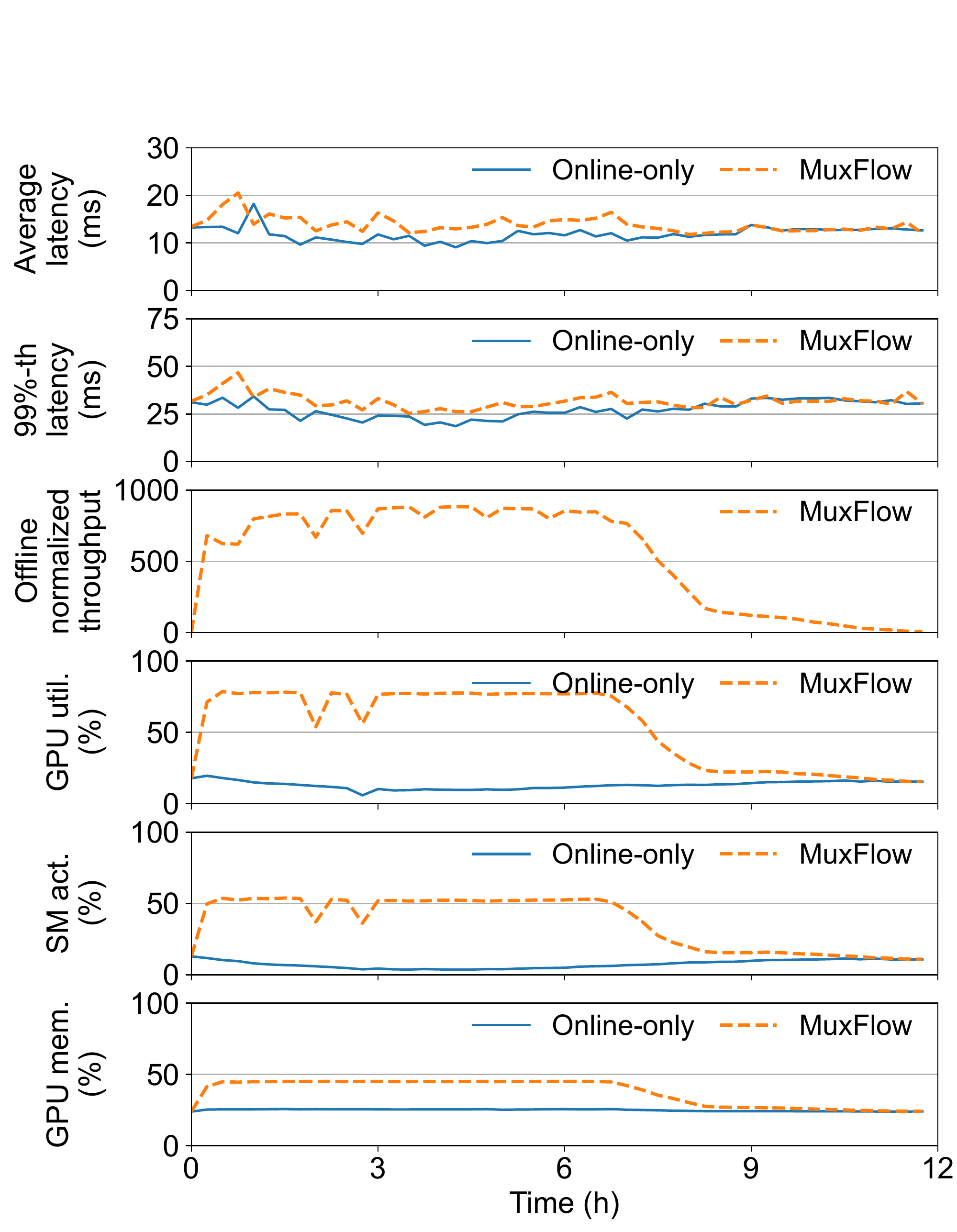}}
        \vspace{-0.15in}
        \caption{Detailed metrics in the testbed experiment.}
        \vspace{-0.1in}
        \label{fig:eval_testbed}
\end{figure}

We first evaluate \sysname on a testbed with 1,000 GPUs.
Figure~\ref{fig:eval_testbed} shows the detailed metrics for online workloads, offline workloads, and GPU resource utilization.
To get the metrics of Online-only, we stop the offline workloads for one minute in every scheduling interval and collect the metrics.
The scheduling interval is set to 15 minutes considering the overhead of pulling images and initialization.
\revise{
Though the trace of offline workloads lasts for 12 hours, most workloads finish before 8 hours.
Thus, there is an obvious shift for \sysname's curves of offline normalized throughput and GPU resource utilizations between 7 to 8 hours.
}

\begin{figure*}[t]
	\subfigure[Latency of online workloads.]{
        \begin{minipage}{0.32\linewidth}
        \centerline{\includegraphics[width=\linewidth,trim=10 0 20 0,clip]{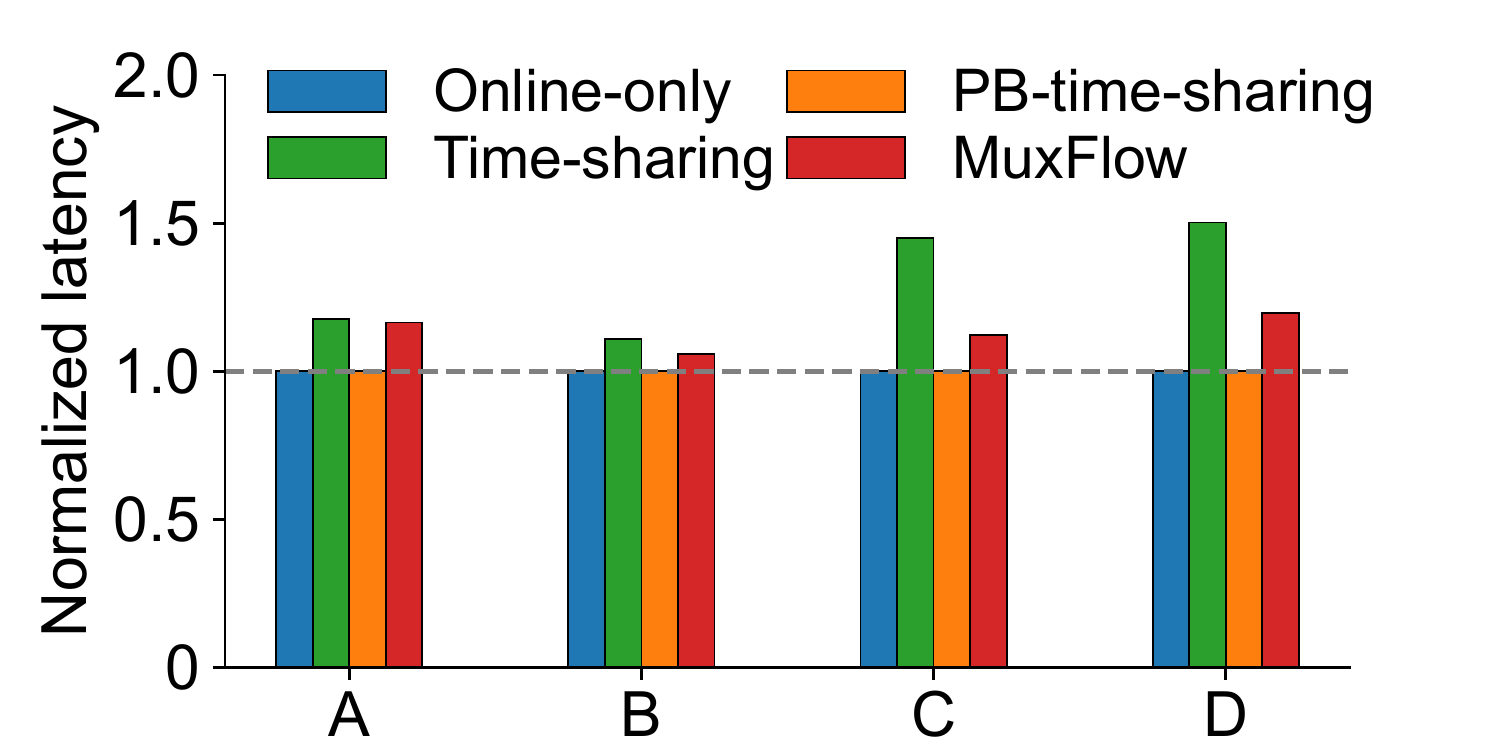}}
        \label{fig:eval_baseline_avglatency}
        \end{minipage}
    }
	\subfigure[JCT of offline workloads.]{
        \begin{minipage}{0.32\linewidth}
        \centerline{\includegraphics[width=\linewidth,trim=0 0 20 0,clip]{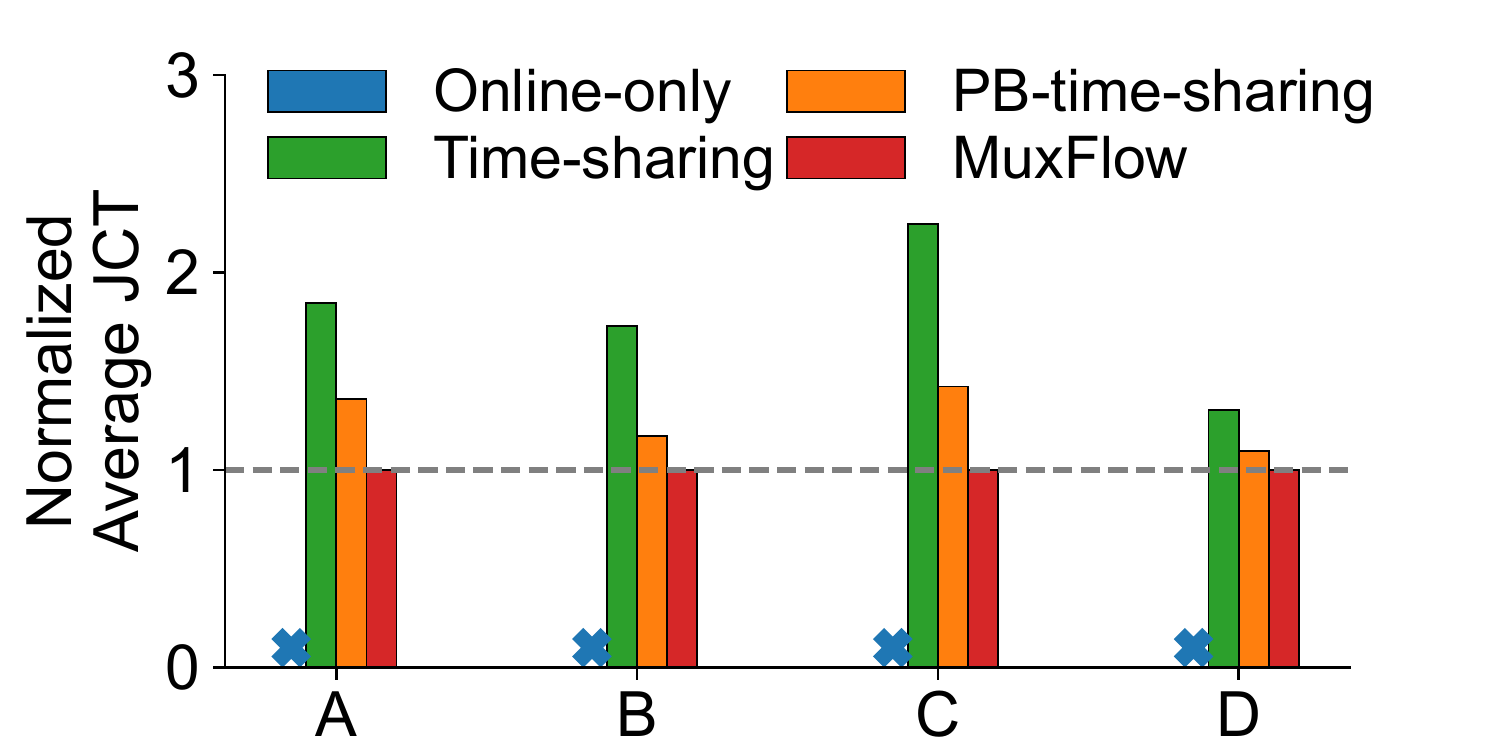}}
        \label{fig:eval_baseline_jct}
        \end{minipage}
    }
	\subfigure[Oversold GPU.]{
        \begin{minipage}{0.32\linewidth}
        \centerline{\includegraphics[width=\linewidth,trim=10 0 20 0,clip]{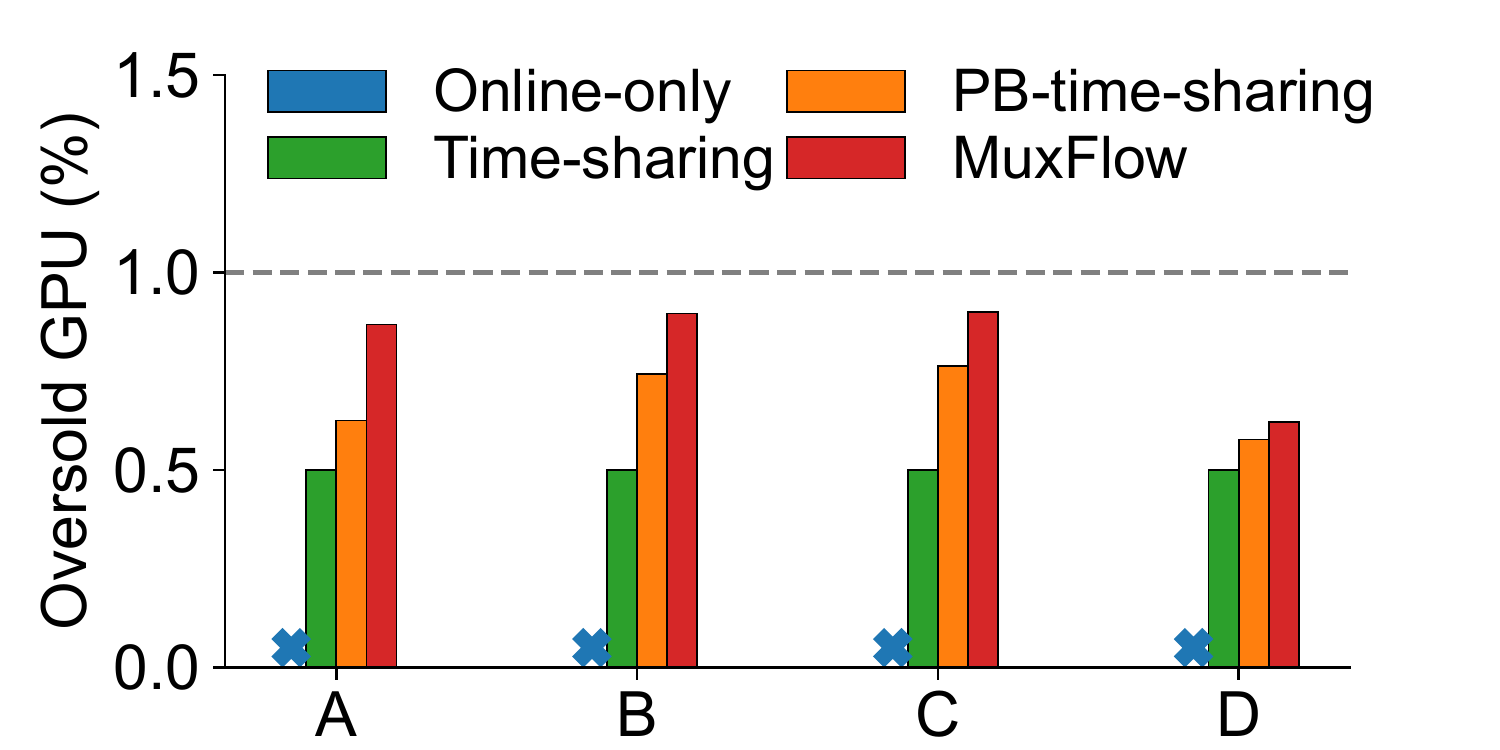}}
        \label{fig:eval_baseline_oversold}
        \end{minipage}
    }
     \vspace{-0.2in}
     \caption{Compare \sysname with related work.}
     \vspace{-0.1in}
    \label{fig:eval_baseline}
\end{figure*}

\parabf{Performance of online workloads.}
Figure~\ref{fig:eval_testbed} shows that \sysname increases the average latency by $16.0\%$, the $99\%-th$ latency by $15.3\%$.
These results indicate that \sysname slows down online workloads less than $20\%$, i.e., 10ms.
It is worth mentioning that such a slowdown is almost imperceptible for most online workloads, e.g. recommendation services and machine translation.
In the wide spectrum of industrial online workloads deployed in  \company, the latency demand of most online workloads is more than 100ms, hence the 10ms slowdown of online workloads is acceptable in practice. 
Additionally, we can adjust \bytecuda and the dynamic SM allocation mechanism to reduce the slowdown threshold or improve the oversold resource for offline workloads.
\revise{
We observe that $1.5\%$ executions of offline workloads are evicted, indicating the functionality of performance protection mechanisms.
}

\parabf{Efficiency of offline workloads.}
We find that \sysname provides up to $86.42\%$ GPU resource to offline workloads, which is a substantial number considering the large number of GPUs in production.

\parabf{GPU resource utilization.}
Figure~\ref{fig:eval_testbed} compares the GPU computing utilization and memory usage between Online-only and \sysname.
The utilization numbers are the average of all GPUs.
\sysname improves the GPU utilization by $4.0\times$, SM activity by $4.7\times$, and GPU memory usage by $1.5\times$.

\parabf{Safety.}
We observe that during the 12-hour testbed experiments, no error propagation happens.
That is, no online workload is influenced by offline workload errors, verifying the safety of \sysname.

\subsection{Comparison with related work}

We compare \sysname with three related systems, Online-only, Time-sharing, and PB-time-sharing.
The related systems are implemented in our simulator, and evaluated with production online workloads and popular offline workloads.
Figure~\ref{fig:eval_baseline} demonstrates the average latency of online workloads, average JCT of offline workloads, and oversold GPU to offline workloads.
We normalize the latency by that of Online-only and other metrics by that of \sysname.
\sysname improves the average JCT by $1.10-2.24\times$, and the oversold GPU by $1.08-1.97\times$, while slowing down the online workloads by less than $20\%$.
Time-sharing slows down online workloads by up to $50\%$, indicating a great impact on online workloads.
PB-time-sharing utilizes priority to protect the performance of online workloads, but its metrics for offline workloads are worse than \sysname due to two reasons.
First, \sysname can utilize the GPU resource wasted by online workloads in space.
Second, \sysname employs the scheduling algorithm to improve the efficiency of offline workloads.

\subsection{Analysis of \sysname}

\begin{figure}[t]
	\subfigure[Impact of the hidden size.]{
        \begin{minipage}{0.45\linewidth}
        \centerline{\includegraphics[width=\linewidth]{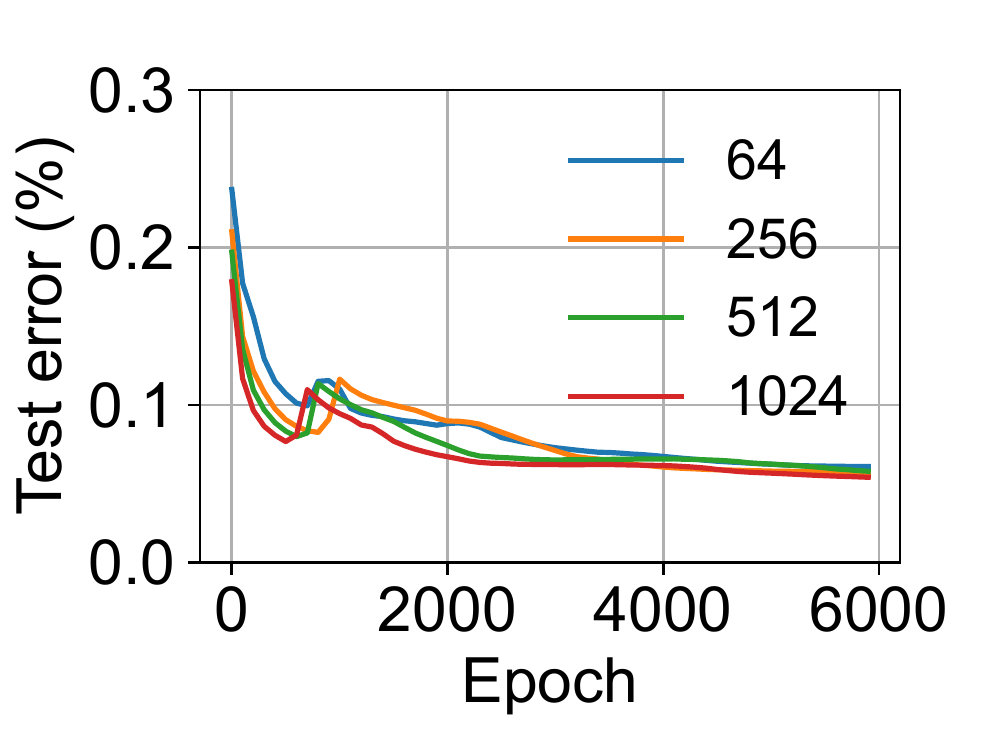}}
        \label{fig:eval_mlp_hidden_size}
        \end{minipage}
    }
	\subfigure[Impact of the number of layers.]{
        \begin{minipage}{0.45\linewidth}
        \centerline{\includegraphics[width=\linewidth]{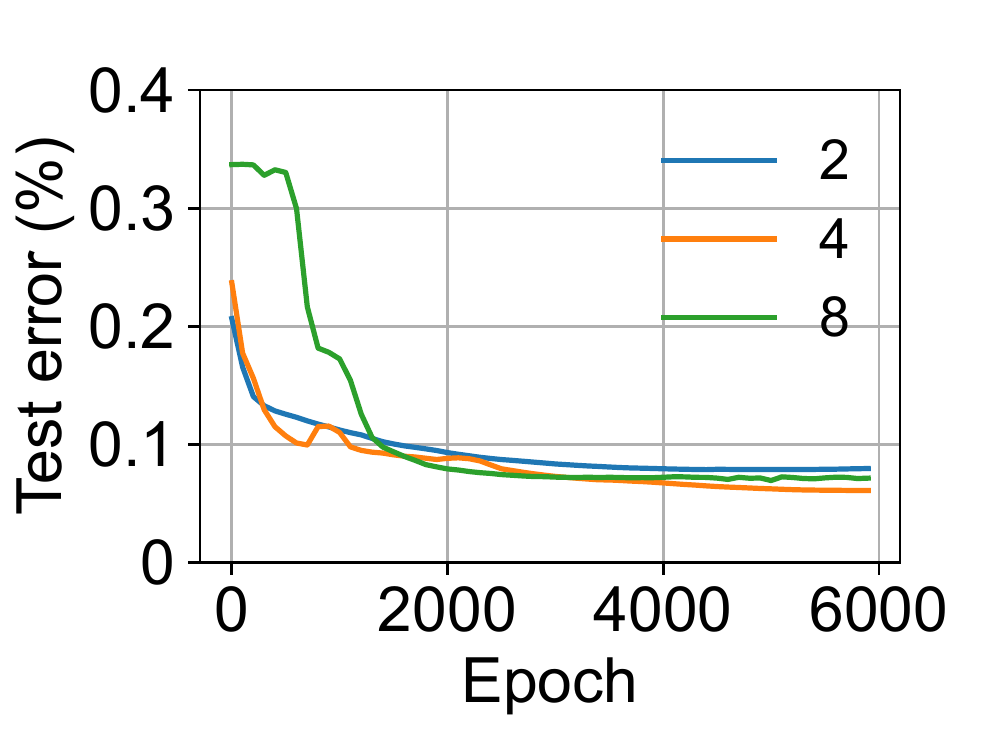}}
        \label{fig:eval_mlp_layers}
        \end{minipage}
    }
     \vspace{-0.15in}
     \caption{Impact of the MLP architecture on the prediction accuracy.}
     \vspace{-0.1in}
    \label{fig:eval_mlp}
\end{figure}

\parabf{Accuracy of the speed predictor.}
\revise{
To better understand the impact of MLP architecture on prediction accuracy, we evaluate the speed predictor with various hidden sizes and numbers of the network layers in MLP.
}
We vary the hidden size from $64\times 64$ to $1024\times 1024$ and fix the number of layers to 4.
The test error curves in Figure~\ref{fig:eval_mlp_hidden_size} show that the MLPs with different hidden sizes have similar accuracy and convergence speed.
For the number of layers, we evaluate the MLP with 2 to 8 layers as shown in Figure~\ref{fig:eval_mlp_layers}, with the hidden size fixed to $64\times 64$.
We find that the error is lowest for 4 layers due to its proper relationship between dataset size and parameters.
Thus, we select $64\times 64$ as the hidden size and $4$ as the layer number for better accuracy and faster inference time.

\begin{figure*}[t]
	\subfigure[Latency of online workloads.]{
        \begin{minipage}{0.32\linewidth}
        \centerline{\includegraphics[width=\linewidth,trim=10 0 20 0,clip]{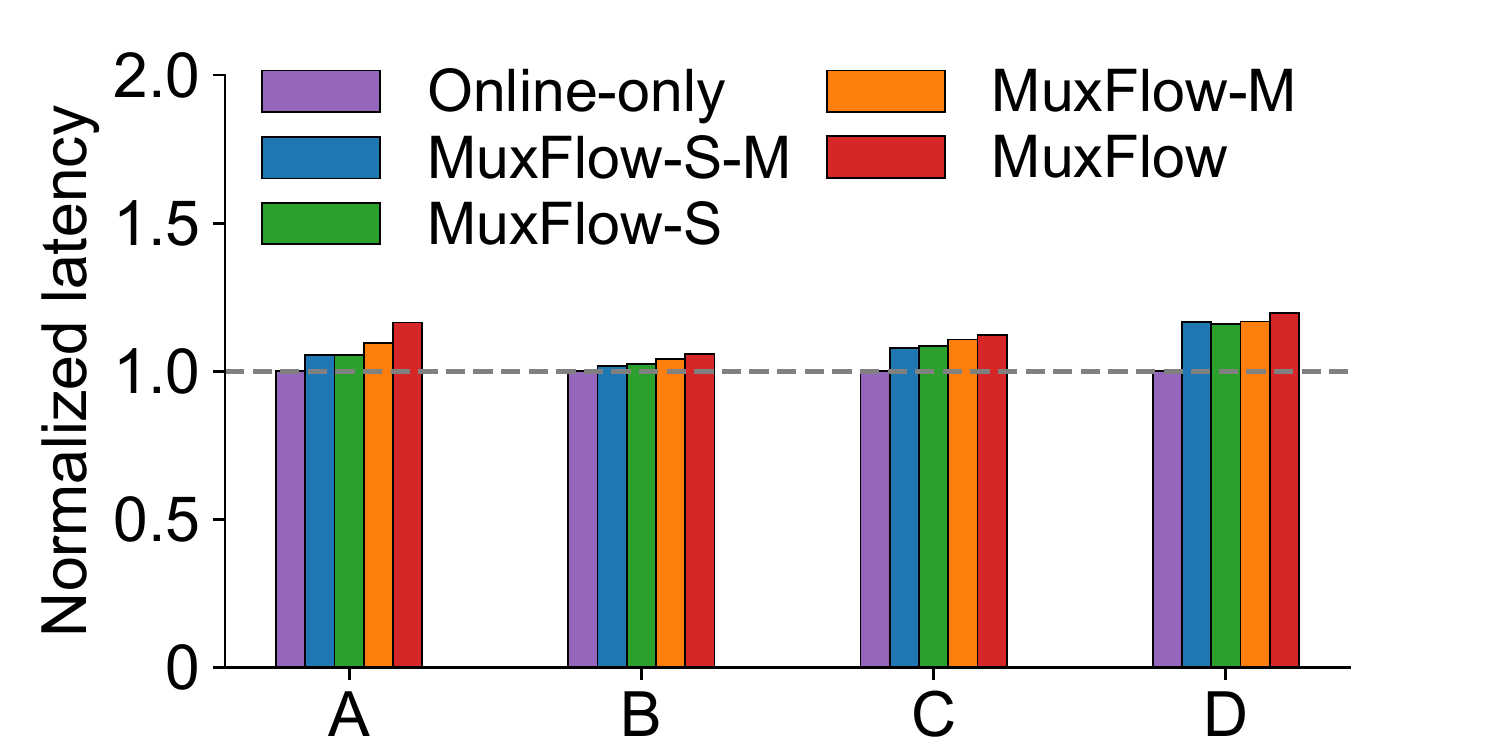}}
        \label{fig:eval_design_avglatency}
        \end{minipage}
    }
	\subfigure[JCT of offline workloads.]{
        \begin{minipage}{0.32\linewidth}
        \centerline{\includegraphics[width=\linewidth,trim=0 0 20 0,clip]{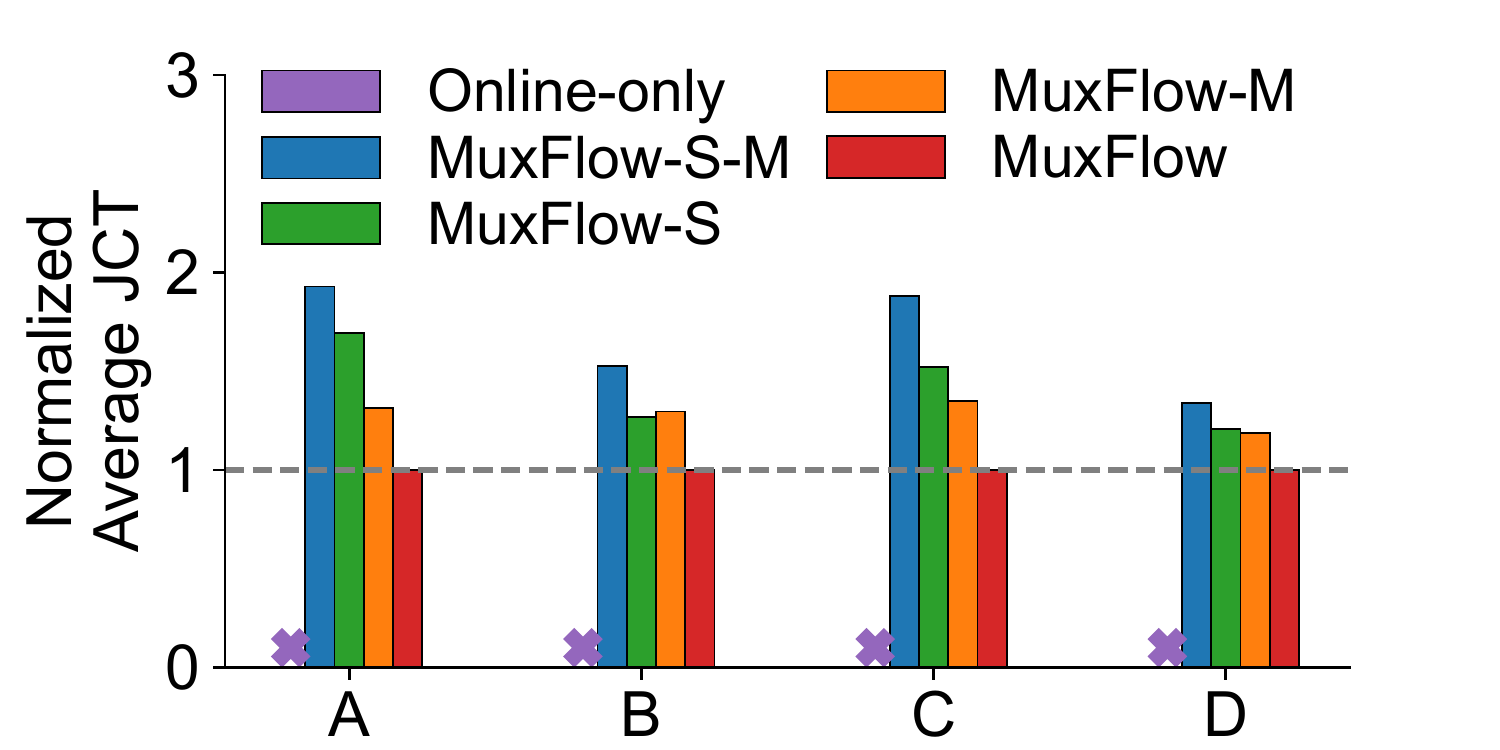}}
        \label{fig:eval_design_jct}
        \end{minipage}
    }
	\subfigure[Oversold GPU.]{
        \begin{minipage}{0.32\linewidth}
        \centerline{\includegraphics[width=\linewidth,trim=10 0 20 0,clip]{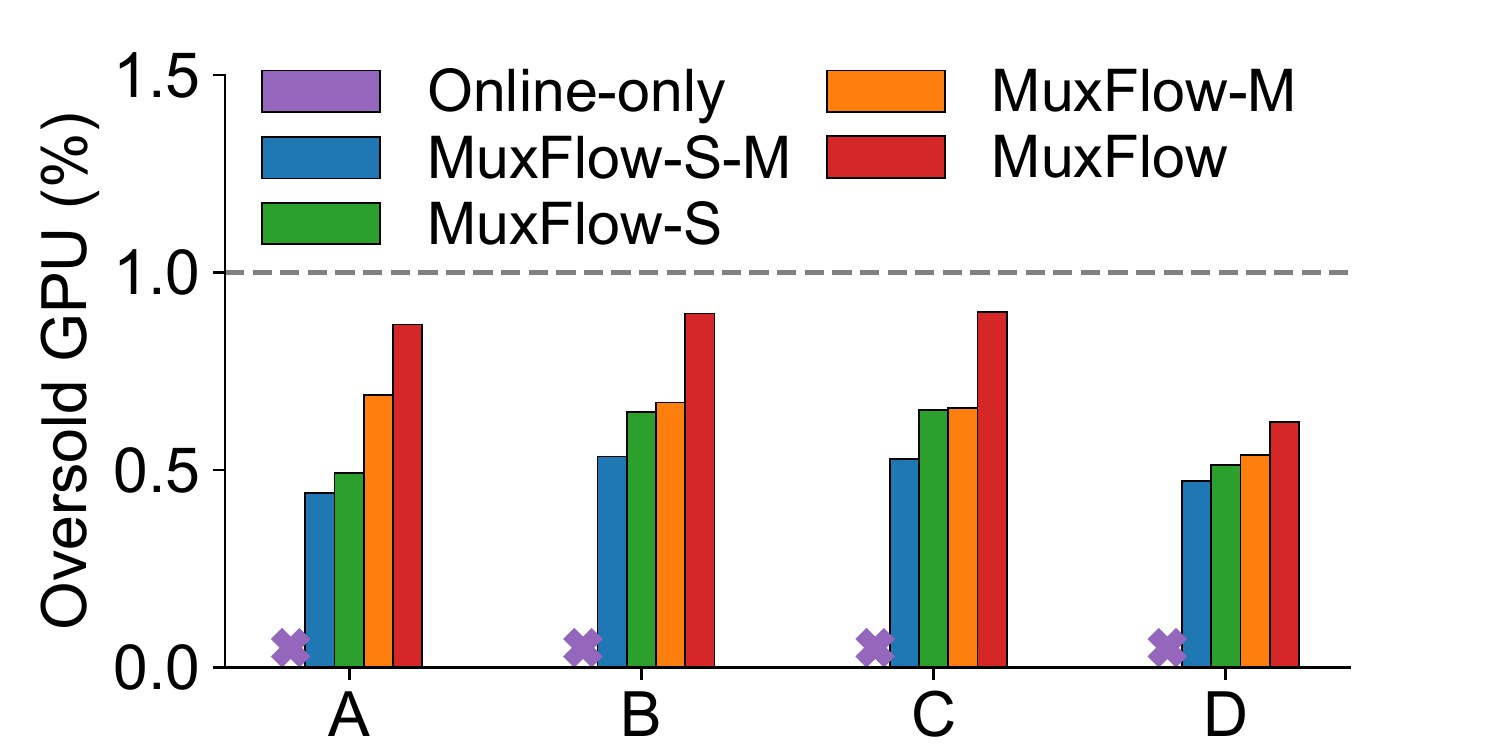}}
        \label{fig:eval_design_oversold}
        \end{minipage}
    }
     \vspace{-0.2in}
     \caption{Impact of the mechanisms for offline efficiency.}
     \vspace{-0.1in}
    \label{fig:eval_design}
\end{figure*}
\parabf{Impact of the mechanisms for offline efficiency.}
We also study the impact of the dynamic SM allocation mechanism and the matching-based scheduling.
We leverage the simulator to compare \sysname with three variants, \sysname without dynamic SM allocation mechanism (\sysname-S), \sysname without matching-based scheduling (\sysname-M), 
\revise{
and \sysname without both dynamic SM allocation mechanism and matching-based scheduling but only the protection mechanisms for online workloads  (\sysname-S-M).
}
Figure~\ref{fig:eval_design} reports the metrics for online and offline workloads over traces A to D.
We normalize the latency by that of Online-only and average JCT by that of \sysname.
Note that compared with \sysname-S-M, both \sysname-S and \sysname-M improve the average JCT and oversold GPU.
\revise{
These improvements confirm that only the online protection mechanisms are not efficient, and dynamic SM allocation mechanism and matching-based scheduling are important for offline efficiency.
}
Additionally, combining the two mechanisms, i.e., \sysname, brings more benefits.

\parabf{System overhead.}
\revise{
\sysname mainly introduces two kinds of overhead, i.e., the profiling overhead and the scheduling overhead.
The profiling takes less than 10 minutes for each offline workload.
The profiling overhead is marginal, as the offline workloads usually take hours or even days,
The overhead of the scheduling algorithm consists of two periods.
The first period is to predict the sharing performance and build the bipartite graph.
Each prediction only takes less than one millisecond, and each internal cluster at \company consists of thousands of GPUs and thousands of workloads.
Thus, the first period only takes several seconds for each cluster using the batched prediction.
The second period is to execute the KM algorithm, which takes several minutes for thousands of workloads.
Note that the scheduling algorithm can be executed in parallel with the workload execution.
Thus, we can hide the scheduling overhead within each scheduling interval.
}

\subsection{Production deployment}

\begin{figure}[t]
	\subfigure[Latency.]{
        \begin{minipage}{0.47\linewidth}
        \centerline{\includegraphics[width=\linewidth,trim=0 0 0 20,clip]{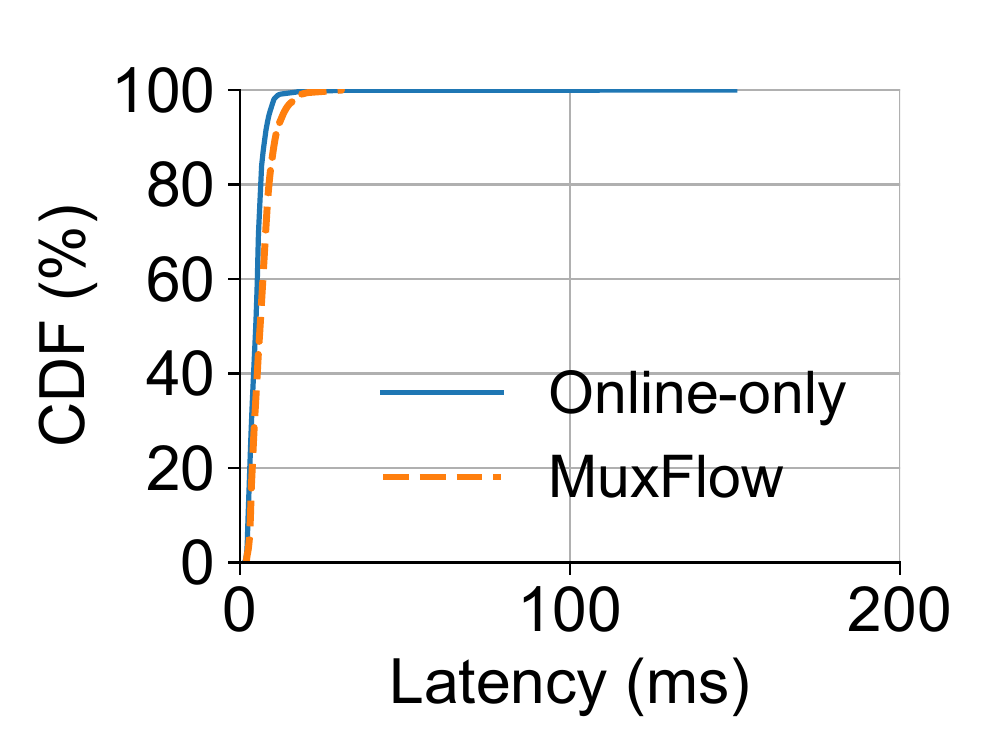}}
        \label{fig:eval_deploy_latency}
        \end{minipage}
    }
	\subfigure[Throughput.]{
        \begin{minipage}{0.47\linewidth}
        \centerline{\includegraphics[width=\linewidth,trim=0 0 0 20,clip]{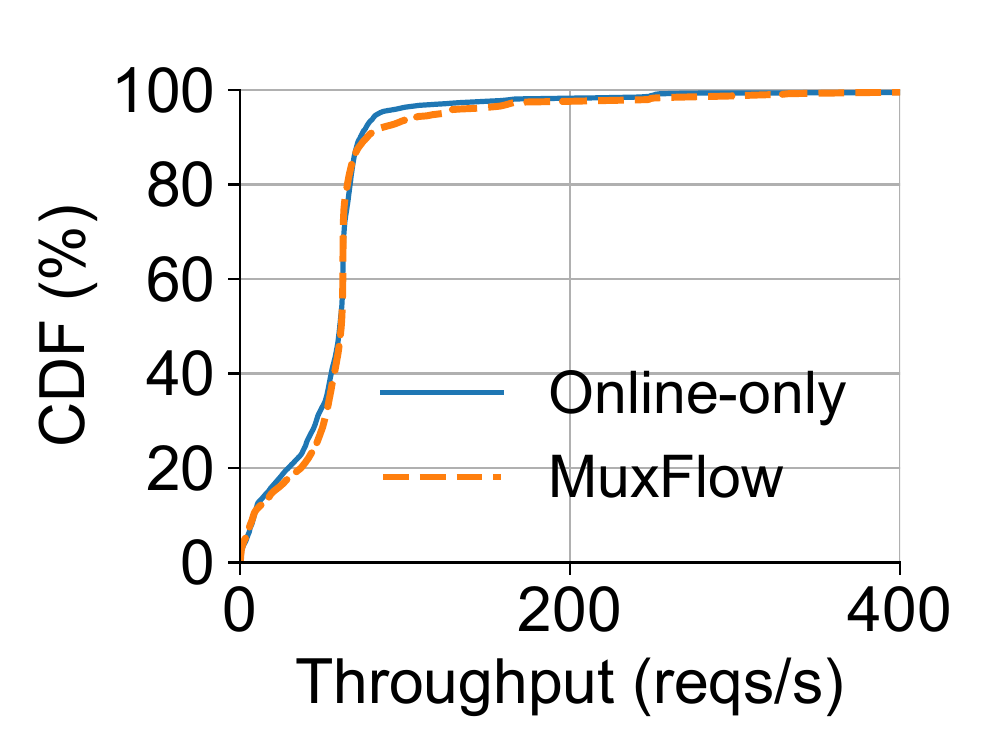}}
        \label{fig:eval_deploy_qps}
        \end{minipage}
    }
     \vspace{-0.15in}
     \caption{Performance of online workloads in deployment.}
    \label{fig:eval_deploy_online}
\end{figure}

\begin{figure}[t]
    \centerline{\includegraphics[width=\linewidth,trim=0 0 0 30, clip]{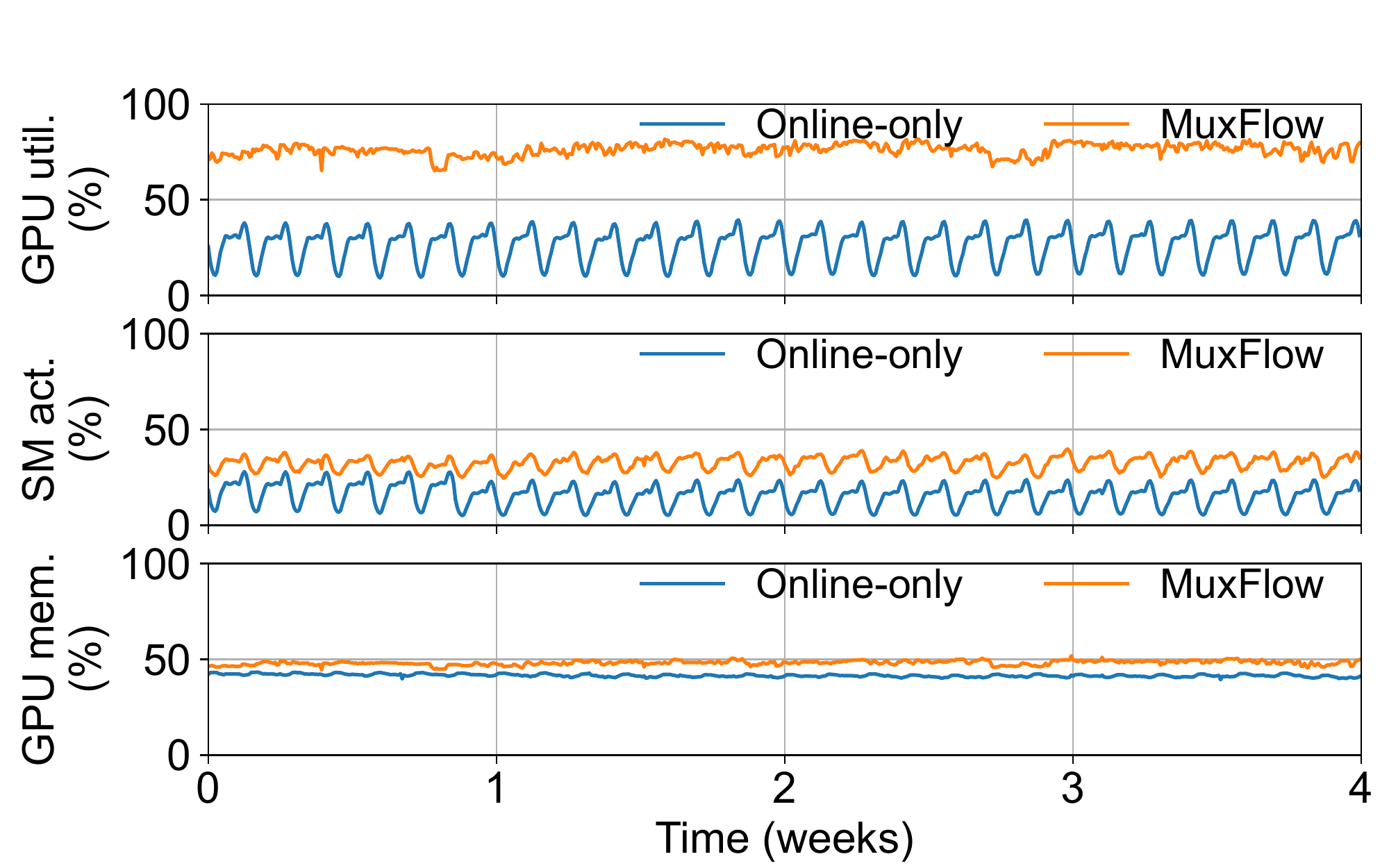}}
    \vspace{-0.15in}
    \caption{GPU resource utilization in deployment.}
    \vspace{-0.1in}
    \label{fig:eval_deploy_gpu}
\end{figure}

We have deployed \sysname on the production clusters with more than 20,000 GPUs at \company.
As the results of the whole system are not ready when the paper is written, we concentrate on the results of \sysname without the dynamic SM allocation mechanism and the matching-based scheduling.
\revise{
To verify the performance protection for online workloads provided by \sysname, we collect the latency and throughput of online workloads that are deployed with both \sysname and dedicated inference clusters (Online-only), as shown in Figure~\ref{fig:eval_deploy_online}.
The average latency and $99\%-th$ latency of online workloads increase by less than $10ms$.
The slowdown of online workloads is acceptable compared with the latency demand of our online workloads.
Besides, we collect the average resource utilization of all GPUs used by \sysname and Online-only for four weeks in Figure~\ref{fig:eval_deploy_gpu}.
We observe that \sysname improves the GPU utilization from $26\%$ to $76\%$, SM activity from $16\%$ to $33\%$, and GPU memory from $42\%$ to $48\%$, indicating the efficiency of \sysname.
The GPU memory utilization increases less than other utilizations because of our conservative memory limitation for offline workloads.
}

\revise{
The percentage of daily error devices of \sysname is below $0.9\%$, which is slightly higher than the error rate of Online-only at \company, $0.7\%$.
However, the increase of the error rate is much less than the increase of the executed containers, i.e., $2\times$.
Compared with Online-only, the extra device errors of \sysname come from MPS server crashes and other MPS hangs, which cannot be handled with existing mechanisms.
}

%% file: sections/experience.tex
\section{Lessons and future direction}
\label{sec:experience}

\parabf{Safety protection.}
How to guarantee the safety of shared workloads is one of the most important problems of deploying GPU sharing.
Thus, most GPU sharing solutions~\cite{cgpu,gu2018gaiagpu} in production do not consider MPS due to its poor isolation ability.
In contrast, to our best knowledge, we are the first to thoroughly analyze all unsafe cases encountered in our production cluster and propose corresponding solutions for these cases.
Almost all unsafe cases in our deployment can be handled by the graceful exit mechanism of \sysname.
Besides, we have worked with NVIDIA to improve MPS.
Some bugs and features we reported have been verified and fixed by NVIDIA.
For example, we observed that sharing workloads compiled by different GCC versions with MPS can cause the MPS server hangs, and this problem has been verified by NVIDIA and fixed in NVIDIA GPU Driver 470.

However, things become complicated when considering malicious behaviors, e.g., intriguing sticky CUDA error by dividing zero to influence the shared workload.
To avoid malicious behaviors, \sysname only accepts trustworthy offline workloads and we employ a fault detector with manually-summarized rules to monitor collected device metrics and alert when abnormal situations are found.
\revise{
For now, \sysname is only used in our internal clusters and for internal users. 
These protection approaches seem safe enough according to our deployment experience.
Yet if considering external users in a cloud setting,
we need more general and automated approaches for safety protection and malicious behaviour detection.
One opportunity is to leverage DL to discover malicious behaviours automatically~\cite{saufi2019challenges}.
Besides, enabling the scheduler to identify fail-prone workloads and avoid sharing them with other workloads is another possible approach.
}

\parabf{Slowdown of online workloads.}
In this paper, we get the latency of online workloads increases less than $20\%$, i.e., $10ms$.
The degradation is affordable and acceptable in our internal cluster, because most latency demands are more than 100ms for production online workloads.
Note that the degradation threshold is a tradeoff between the online service quality and resource utilization, and it can be changed in \sysname by two mechanisms.
First, the parameters of GPU load~\ref{equ:gpu_load}$\&$\ref{equ:clock_factor} in \bytecuda affect how the offline workload is executed and then how the online workload is influenced.
Second, we can adjust the SM percentage assigned to offline workloads to change the slowdown of online workloads.
How to select a proper degradation threshold for each cluster or even each online workload is left as an open problem.

\parabf{The number of shared workloads.}
In \sysname, we share at most one offline workload with each online workload because one offline workload is usually enough to fill SMs up.
Sharing multiple offline workloads with multiple online workloads may bring more benefits, especially for light-weighted offline workloads.
However, there are four challenges to sharing multiple workloads.
First, we need to guarantee the performance of all online workloads.
Second, we need to limit the total SM percentage used by multiple offline workloads which cannot be simply limited by MPS parameters.
Third, \bytecuda needs to monitor kernel launches of all offline workloads and decide which kernel to delay or launch according to their priority.
Fourth, the scheduling algorithm to choose sharing pairs with more than three workloads becomes an NP-hard problem~\cite{zhao2022multi}.
\revise{
How to solve these challenges to utilize GPU better is an interesting future direction.
}

%% file: sections/related.tex
\section{Related work}
\label{sec:related}

\paraf{DL workload scheduling.}
Existing DL schedulers are mainly designed for online or offline workloads, but not both, to ensure the performance of online workloads and avoid interference.
The primary goals of online workload schedulers~\cite{crankshaw2017clipper,shen2019nexus,gujarati2020serving,romero2021infaas} are meeting the latency demand and improving overall throughput.
However, these schedulers let one workload monopolize GPUs and thus, cannot fully exploit the GPU resource.
Most prior offline workload schedulers~\cite{gu2019tiresias,hwang2021elastic,qiao2020pollux,mohan2022synergy} also allocate GPUs exclusively.
However, existing offline workload schedulers cannot be directly applied to GPU-sharing clusters because they cannot ensure the performance of high-priority online workloads.
Differently, \sysname leverages space-sharing to improve GPU resource utilization and applies a two-level protection mechanism to guarantee the performance of online workloads.

\parabf{Resource sharing for big data workloads.}
Prior work has studied resource sharing for big data workloads and CPU clusters.
DRF~\cite{ghodsi2011dominant} extends max-min fairness to achieve resource sharing fairness.
Tetris~\cite{grandl2014multi}, Graphene~\cite{grandl2016graphene}, and Carbyne~\cite{grandl2016altruistic} propose heuristic algorithms to solve multi-resource scheduling problems and improve resource utilization.
MonoSpark~\cite{ousterhout2017monotasks} improves performance clarity by splitting data analytics tasks into monotasks.
Many large enterprises deploy resource-sharing clusters.
Apollo system~\cite{boutin2014apollo} improves resource utilization by opportunistic tasks in Microsoft.
Google's Borg~\cite{verma2015large,tirmazi2020borg} adopts machine sharing with performance isolation to achieve high utilization.
In comparison, DL workloads use GPUs to speed up, and GPUs lack efficient and safe sharing mechanisms.
Thus, it is more challenging to deploy GPU sharing in production clusters.

\parabf{GPU sharing for DL workloads.}
Recently, GPU sharing has been studied for DL workloads.
The techniques to share GPUs mainly fall into two categories.
Prior time-sharing approaches~\cite{xiao2018gandiva,wang2021wavelet,lim2021zico,zhao2022multi} may impact the efficiency of shared workloads and cannot fully utilize GPU computing power in every time slice.
\revise{
Salus~\cite{yu2019salus} and PipeSwitch~\cite{bai2020pipeswitch} propose fast job switching and memory management techniques to speed up time-sharing.
But they cannot avoid the intrinsic drawbacks of time-sharing.
}
Some work~\cite{gu2018gaiagpu,xiao2020antman,weng2022pai,cgpu} assigns time slices according to priority to guarantee the performance of online workloads.
However, these approaches still cannot improve resource utilization for each time slice.
\sysname employs space-sharing and performance protection mechanisms for efficient and safe GPU sharing.

Space-sharing is the other direction to share GPU.
NVIDIA's MPS~\cite{mps} is a general method to multiplex jobs on NVIDIA GPUs.
Gavel~\cite{narayanan2020heterogeneity} leverages MPS directly but it cannot guarantee the performance of online workloads.
\revise{
GSLICE~\cite{dhakal2020gslice} advances MPS to support dynamic and fair resource allocation but it does not consider cluster-level scheduling.
}
Retiarii~\cite{zhang2020retiarii} merges multiple similar models to improve GPU utilization which is infeasible for production clusters running diverse workloads.
\revise{
DeepPool~\cite{park2022efficient} and Reef~\cite{han2022microsecond} leverage priority-based multi-stream approaches.
However, multi-stream approaches are unfit for production deployment mainly due to two reasons.
First, it needs to manually merge different workloads into one process to leverage NVIDIA GPU streams, which is not friendly to existing infrastructure and users.
Second, multiple streams may introduce the overhead of locks and CPU kernel launches.
}
In contrast, \sysname is a practical space-sharing cluster system that has been deployed at \company.

\revise{
Some work predicts performance interference among shared workloads for time-sharing~\cite{chen2016baymax,chen2017prophet} and space-sharing~\cite{zhao2019themis}.
These approaches play the similar role as the DL-based speed predictor in \sysname, and are orthogonal to other system designs.
}

%% file: sections/conclusion.tex
\section{Conclusion}
\label{sec:conclusion}

In this paper, we have presented \sysname, the first production DL cluster system for efficient and safe space-sharing.
\sysname ensures the performance of online workloads from both workload level and GPU level.
To guarantee the safety of online workloads, \sysname leverages a mixed error-handling mechanism based on the analysis of production errors.
Furthermore, \sysname exploits dynamical SM allocation and matching-based scheduling to improve the efficiency of offline workloads.
The evaluation results demonstrate the efficiency and efficacy of \sysname.
Particularly, \sysname has been already deployed in the production DL clusters at \company with more than 20,000 GPUs.